\documentclass[aps,pre,reprint,showpacs,twocolumn,superscriptaddress,usenames,dvipsnames]{revtex4-1}
\usepackage{amsmath}
\usepackage{amssymb}
\usepackage{graphicx}% Include figure files
\usepackage{dcolumn}% Align table columns on decimal point
\usepackage{bm}% bold math
\usepackage{xcolor}

\newcommand{\bi}[1]{Fig.~\ref{fig:#1}}
\newcommand{\lr}[1]{\left\langle #1 \right\rangle}
\newcommand{\e}[1]{eq.~(\ref{eq:#1})}
\newcommand{\se}[1]{sec.~(\ref{sec:#1})}
\newcommand{\be}{\begin{equation}}
\newcommand{\ee}{\end{equation}}
\newcommand{\ba}{\begin{eqnarray}} 
\newcommand{\ea}{\end{eqnarray}}
\usepackage{color}

\def\mean#1{\langle#1\rangle}

\newcommand{\bracket}[1]{\left(#1\right)}

\begin{document}
\title{Demixing of two species via reciprocally concentration-dependent diffusivity} 

\author{Lutz Schimansky-Geier}
\affiliation{Institute of Physics, Humboldt University at Berlin, Newtonstr. 15, D-12489 Berlin, Germany}

\author{Benjamin Lindner}
\thanks{Corresponding author}
\email{benjamin.lindner@physik.hu-berlin.de}
\affiliation{Bernstein Center for Computational Neuroscience Berlin, Philippstr.~13, Haus 2, 10115 Berlin, Germany}
\affiliation{Institute of Physics, Humboldt University at Berlin, Newtonstr. 15, D-12489 Berlin, Germany}

\author{Sebastian Milster}
\affiliation{Institute of Physics, Humboldt University at Berlin, Newtonstr. 15, D-12489 Berlin, Germany}
\affiliation{Institute of Physics, Albert Ludwig University of Freiburg
Hermann-Herder-Str.~3, D-79104 Freiburg, Germany}

\author{Alexander B. Neiman}
\affiliation{Department of Physics and Astronomy, Ohio University, Athens, Ohio 45701, USA}
\affiliation{Neuroscience Program, Ohio University, Athens, Ohio 45701, USA}
	
\date{\today}
\begin{abstract}
  {We propose a model for demixing of two species by assuming a density-dependent  effective diffusion coefficient of the particles. Both sorts of microswimmers diffuse as active overdamped Brownian particles with a noise intensity that is determined by the surrounding density of the respective other species within a sensing radius $r_s$. A higher concentration of the first (second) sort will enlarge the diffusion and, in consequence, the intensity of the noise experienced by the second (first) sort. Numerical and analytical investigations of steady states of the macroscopic equations prove the demixing of particles due to this reciprocally concentration-dependent diffusivity. An ambiguity of the numerical integration scheme 
for the purely local model ($r_s\to 0$) is resolved by considering nonvanishing sensing radii in a nonlocal model with $r_s > 0$.
}
\end{abstract}
\maketitle
\section{Introduction}
\noindent
Suspensions of mobile active particles are well known to exhibit
various spatio-temporal structures \cite{o2017oscillators,romanczuk2012active,marchetti2013,costanzo2014motility,grossmann2014vortex}  which depend both on the kind of single-particle activity and on the interactions between the particles.  Best
investigated are groups of self-moving units with aligning
interaction as swarms or flocks of animals \cite{vicsek1995novel,gregoire2004onset,couzin2005effective}. Another example are dense bacterial suspensions \cite{volfson2008biomechanical} in which propulsive agents might shape diverse complex flows ranging from compact laminar streaming to turbulent-like patterns \cite{dunkel2013fluid,grossmann2014vortex}. In the recent past, interesting phenomena, as for instance motility-induced phase separation, have been reported for run-and-tumble
bacteria  \cite{tailleur2008statistical}, for self-
propelled Brownian particles \cite{cates2013active} and within the frame of a Cahn Hilliard theory \cite{speck2014effective}. In all these cases, the directed motion of particles creates an self-amplifying instability by increasing the density due to impacts in direction of motion whereby at the backside of the particle the density is depleted  \cite{speck2015dynamical}. This situation was experimental verified for carbon-coated  Janusz-particles \cite{buttinoni2013dynamical} the self-propelled motion of which is based on diffusion phoresis. Comprehensive reviews \cite{cates2015motility,bergmann2018active} summarize the phase-separating process of the particles.

These studies have inspired a larger  number of investigations on phase separation in suspensions of self-moving object. Various interesting problems have been put forward as for example, the mixtures of active and passive or fast and slow particles \cite{grosberg2015nonequilibrium,stenhammar2015activity,takatori2015theory,
tanaka2017hot,sese2018velocity,rodriguez2020phase,kolb2020active}, the influences of different speeds of phase separation \cite{baglietto2020otherwise}, different diffusivities \cite{weber2016binary}, a discontinuous motility \cite{elliott1996cahn,fischer2020quorum}, demixing of active particles in external fields \cite{kumari2017demixing}, chiral active matter \cite{liebchen2017collective}, descriptions of phase separation far from equilibrium in continuum frame  \cite{stenhammar2013continuum}, learning of groups in swarms \cite{durve2020learning} and chemotactically reacting particles \cite{meyer2014active,rapp2019universal}, to mention just a few publications. Also early work on chemically reacting active Brownian particles creating complex behavior during trail formation of  ants and in excitable dynamics deserves to be mentioned \cite{schweitzer1994clustering,schimansky1995structure}. Especially in biophysical applications studies of phase separation  might gain importance in the long term, see for example \cite{liu2013phase,st2010cell,palacci2013living,murray2017self}.

Here we put forward a minimalistic model for a demixing interaction of two species of diffusing particles in which the effective diffusion coefficient of one sort (say sort $A$) is controlled by the neighboring probability density function (pdf)  $p_B$ of the particles of the respective other sort (say sort $B$). In this model, the diffusion of the $A$ particles is increased by a factor that depends on the power of the density $p_B^q$, $q \,\in \, \mathbb{N}$,  with which this enters into the model. As we will show, using $q=1$ (a linear dependence of the diffusion coefficient on the density) does not result in a demixing despite the nonlinear character of the corresponding macroscopic fluxes. However, the most simple nonlinear dependence, namely a quadratic dependence ($q=2$), entering with a sufficiently high prefactor
causes a demixing of the two species.  

Demixing will be demonstrated by particle simulations, by a stability analysis and by numerical integration of the asymptotic macroscopic equations for $p_A$ and $p_B$. A technically challenging but interesting issue is how the pdf used in the dynamics is incorporated. We also vary the value of the sensing radius $r_s$ and demonstrate that also for nonvanishing  radius particle separation for sufficiently strong nonlinearity can be found. In contrast, a purely local sensing results in an ambiguity of the mathematical description. Configurations with steep jumping interfaces and with multiple domains are found.  Inconclusive ambiguous results which depend on time steps as well as on the underlying integration grid are found and do no allow unique answers. Here we study this interesting situation in detail.

The corresponding microscopic dynamics of the two kinds of particles are overdamped Langevin equations. Such dynamics results from an adiabatic elimination of inertia in models for stochastic microswimmers with constant speed $v_0$ and angular noise with intensity $D_{\phi}$ \cite{romanczuk2012active} and describes diffusional motion with the effective diffusion coefficient: 
\begin{equation}
D_{{\rm eff}}\,=\, v_0^4 /(2D_{\phi})
\label{eq:eff_diff}
\end{equation}
(for related results, see \cite{MikMei98,LinNic08b}). Therefore, changes or modifications of the effective diffusion coefficient $D_{{\rm eff}}$ are caused  by an alteration of the propulsive apparatus, i.e. of the speed $v_0$  or the angular noise $D_{\phi}$,

As already mentioned, the strength of the effective self-mobility or of the mobile response of the particles might be also controlled by chemotactic or phoretic forces generated self-consistently by the ensemble
\cite{schweitzer1994clustering,{golestanian2019phoretic},soto2014self,soto2015self,agudo2019active,nasouri2020exact}. In a simple example, particles distribute a chemical substrate which creates a common memory field of the former particle motion. The members of the ensemble respond to the strength of this field by changing their  diffusivities. 
\section{Demixing of active particles: Model}
In this paper, we propose a simple model which exhibits phase separation (demixing) of two types of agents, referred to as $A$ and $B$ in what follows. We assume a symmetry between both species with respect to all parameters. First of all, each population contains the same number $N$ of particles. Asymmetric setups will show similar results but will complicate the problem. Secondly, both $A$ and $B$ particles perform an overdamped Brownian motion \cite{hanggi1982stochastic,chavanis2014brownian} as diffusional approximation of stochastic microswimmers \cite{milster2017eliminating,noetel2017adiabatic}.
Thirdly, the individual diffusion coefficient of, say the $i$th $A$ particle is a functional of the $B$ particles' density $p_B$ taken around the current position $x_{A,i}(t)$ inside a sensing domain with spatial extension $r_s$. Such sensing regions are very popular in investigations of animal motion \cite{couzin2005effective} and swarming models \cite{vicsek1995novel,romanczuk2012active,marchetti2013}. We distinguish between a \emph{local} version of the model, in which we use an estimate of the density of $B$ particles at $x_{A,i}(t)$ and a \emph{non-local} version, in which the density of $B$ particles within the sensing radius $r_s$ enters.

Further on, we will restrict ourselves to an one dimensional setting and denote particle positions  by $x_{A,i}$ and $x_{B,j}$, respectively, where $i,j=1,\ldots, N$. Particles can move in the interval $[-\ell,\ell]$ and we apply periodic or reflecting (no-flux) boundary conditions. 

As the central statistics of interest we consider the long-time asymptotics of the pdfs $p_A(x)$ and $p_B(x)$ but we will also briefly discuss transient behavior.  The densities gain physical meaning if connected with a spatial grid of $N_{bin}$ elements with spatial extension $\Delta x= 2\ell/N_{bin}$. Inside the $\Delta x$ the pdfs are assumed to be constants and hence, the partition of $N_{bin}$ elements defines the accuracy of our output.

We also will scale the sensing radius $r_s$ in units of the introduced bin $\Delta x$. In detail we will set 
\be
r_s = (s-\frac{1}{2})\Delta x, \quad s=1,2,... .
\label{eq:rs-disc}
\ee
 In consequence, for $s>1$ we describe situations in which the pdf can be inhomogeneous within the sensing region.

The dynamics of the particle's positions is given by the set of Langevin equations
\ba
\frac{\rm d}{{\rm d}t}\,x_{A,i}\,&=&\,\sqrt{2D_0[1+c \mean{p_{B}(x_{A,i})}^q_{r_s}]}\, \xi_{A,i}(t)\nonumber\\
\,~~~~\frac{\rm d}{{\rm d}t}\,x_{B,i}\,&=&\,\sqrt{2D_0[1+c \mean{p_{A}(x_{B,i})}^q_{r_s}]}\, \xi_{B,i}(t).
 \label{eq:diff_lang}  
\ea
Here $\xi_{A,i}(t)$ and $\xi_{B,i}(t)$ are  independent Gaussian white noise sources with vanishing mean and $\mean{\xi_{F,i}(t)\xi_{G,j}(t^\prime)}=\delta_{F,G}\delta_{i,j}\delta(t-t^\prime)$ with $F,G\in \{A,B\}$. Despite their simplicity,  our model equations need some further explanation; specifically, the noise intensities of the fluctuating terms require a number of comments. First of all, as we deal with multiplicative noise, we need an interpretation of the stochastic differential equation (see e.g. \cite{risken1996fokker}); here we will use throughout the \emph{Ito interpretation}. Secondly, we note that the probability density is raised to an integer power $q$ that controls the nonlinearity of the interaction; throughout the paper we will study $q=2$. Thirdly, in a simulation with $2N$ particles it is not clear what we mean by $p_{A,B}(x)$ which is needed for the computation of the noise intensity. 

Last but not least, the brackets under the square roots define a spatial average over the sensing radius in our model. The interaction between the $A$ and $B$ particles takes place in the sensing range, only. With $x_{A,i}(t)$ being the position of the $A$ particles the bracket sums the $B$ particles between $[x_{A,i}(t)-r_s, x_{A,i}(t)+r_s]$ and divides by the length of the sensing domain. This local average is then raised to the $q$th power. On the aforementioned spatial lattice, the average  defines a kind of nonlocal interactions for sensing radii with values $s=2,3 \ldots$. The bracket stands for the integral operator with arbitrary function $f(x)$ at position $x$:
\begin{equation}
\mean{f(x)}_{r_s}\,=\, \frac{1}{2r_s} \int_{-r_s}^{r_s} {\rm d}x^\prime f(x+x^\prime) 
\label{eq:sense}
\end{equation}
We mention that if the sensing radius coincidences with the binning length, one obtains the (spatially discretized) local version of the overdamped dynamics
\ba
\frac{\rm d}{{\rm d}t}\,x_{A,i}\,&=&\,\sqrt{2D_0[1+c p_{B}^q(x_{A,i})]}\, \xi_{A,i}(t)\nonumber\\
\,~~~~\frac{\rm d}{{\rm d}t}\,x_{B,i}\,&=&\,\sqrt{2D_0[1+c p_{A}^q(x_{B,i})]}\, \xi_{B,i}(t).
 \label{eq:diff_lang_local}  
\ea
In order to reduce the number of parameters, we may rescale space and time and replace them by non-dimensional counterparts, $x^\prime = x/\ell$ and $t^\prime =t /\tau$;  using then \e{diff_lang} with $\tau=\ell^2/D_{0}$, we see that we get rid of the parameters $D_0$ and $\ell$ which can be both set to unity. Omitting the primes for the ease of notation, we will use nondimensional variables $x$ and $t$ in the following.
 
Our model equations contain probability density functions (pdfs) that are  not known but can be estimated from the positions of the particles.  
To define the usage of the pdfs in the Langevin equations we link them to the position of particles as follows. For a finite particle number and finite binning length $\Delta x$ we introduce empirical probability function densities 
$\Pi_{A,B}(x,\{x_{A,B,j}\}))$. The latter functions are time-dependent normalized histograms on the above mentioned grid  (we recall that we have $N_{bin}$ discrete bins of width $\Delta x=2\ell/N_{bin}$) with central positions $x_n=-\ell+(n-1/2)\Delta x$ ($n=1,\dots,N_{bin}$). To give an example,  $\Pi_{B}(x,\{x_{B,j}\}))$ measures the fraction of $B$ particles in each bin divided by the bin size and returns this normalized fraction for the bin that contains $x$ (the function's first argument). More formally, if the first argument falls into the $n$th bin ($x_n-\Delta x/2<x< x_n+\Delta x/2$), we can write in terms of Heaviside functions $\Theta(\dots)$
\ba
&&\Pi_{A}(x,\{x_{A,j}\})=\frac1{N \Delta x }\sum_{j=1}^{N} \Theta(\Delta x/2-|x_n-x_{A,j}|)\\
&&\Pi_{B}(x,\{x_{B,j}\})=\frac1{N \Delta x }\sum_{j=1}^{N} \Theta(\Delta x/2-|x_n-x_{B,j}|)
\ea
For an appropriate limit $N \to\infty, \Delta x\to 0$, these functions converge to the probability densities as follows
\be
\Pi_{A,B}(x,\{x_{A,B,j}\}) \to p_{A,B}(x,t).
\label{eq:link_pdf}
\ee
Note that both $\Pi_{A,B}(x,\{x_{A,B,j}\})$ and $p_{A,B}(x,t)$ are separately normalized for each particle sort. The normalization condition for the pdf of $A$ density in the continuous case, for instance, is obviously $\int_{-\ell}^\ell dx\; p_A(x,t)=1$; for the histograms, the condition reads:
\be
\sum_{n=1}^{N_{bin}} \Delta x\; \Pi_{A}(x_n,\{x_{A,j}\})=1. 
\ee
The histogram version of the nonlocal average \e{sense} is carried out as follows
\be
\lr{\Pi_{A,B}(x,\{x_{A,B,j}\})}_{r_s}=\sum_{m=1-s}^{s-1} \frac{\Pi_{A,B}(x_{n+m},\{x_{A,B,j}\})}{{2s-1}}.
\ee 
In all particle simulations, we thus simulate the following version of \e{diff_lang}
(setting now $D_0=1$ and $q=2$ according to the discussion above):
\ba
\frac{ d x_{A,i}}{{d}t}&=&\,\sqrt{1+c \lr{\Pi_B(x_{A,i}(t),\{x_{B,j}\})}_{r_s}^2}\, \xi_{A,i}(t)\nonumber\\
\frac{ d x_{B,i}}{{d}t}&=&\,\sqrt{1+c \lr{\Pi_A(x_{B,i}(t),\{x_{A,j}\})}_{r_s}^2}\, \xi_{B,i}(t)\nonumber\\
 \label{eq:langevin_H}  
\ea
We scale the size of sensing radius \e{rs-disc} in units of the introduced bin $\Delta x$. For the local simulations, we use $s=1$, i.e. $\Pi_{A,B}(x_{A,i}(t),\{x_{B,j}\})$ directly instead of $\lr{\Pi_{A,B}(x_{A,i}(t),\{x_{B,j}\})}_{r_s}$.

Let us finally discuss the dependence of the noise intensities on the pdf of the respective other particle sort  in more detail. The effective diffusion coefficients in \e{langevin_H} grow with the density which is an unusual assumption in equilibrium. However, there are a few Monte Calrlo studies \citep{allen1990diffusion} and model calculations \cite{illien2018nonequilibrium} showing such behavior. Another motivation for this choice comes from early work on population dynamics by Shigesada
\cite{shigesada1979spatial,shigesada1980spatial,mimura1980spatial} and on noise in eclogical systems \cite{spagnolo2004noise}. Members of different social groups repel or attract each other, in our case, diffusively which was modeled by an effective state dependent linear diffusion coefficient $D_{\rm eff}=D_0(1+\sum_j c_j p_j$). 
In particular,  as mentioned our model is also inspired by recent  studies by  Golestanian and coworkers \cite{golestanian2019phoretic,soto2014self,soto2015self,agudo2019active,nasouri2020exact} in which ensembles of chemo-phoretic particles were investigated. We regard our model as a strongly simplified version, in which the details of the chemistry are eliminated and are replaced by an effective control of the diffusion coefficient: the particles of the one species accelerate diffusively if particles of the second kind are present in their vicinity.

\section{Particle simulations: The local case, $s=1$}
\label{sec:particel}

We integrated the overdamped Langevin equations \e{langevin_H} in the local version ($s=1$) with state-dependent diffusion coefficients using a simple Euler-Maruyama scheme in the Ito interpretation of the stochastic differential equations. 
Here and in the following we selected $q=2$ as simplest nonlinearity which exhibits demixing. A linear dependence ($q=1$) of the diffusion coefficient on the density of the other particle sort does not show demixing, which is also in line with theoretical predictions (see Sec. \ref{sec:stability}). We tested both reflecting and periodic boundary conditions at $x=\pm \ell=\pm 1$. In both cases we observed  similar macroscopic configurations with one important difference: under periodic boundary conditions the system exhibits an even number of interfaces whereas with reflecting boundaries this number is odd. If not mentioned otherwise, reflecting boundary conditions are used.

\begin{figure}[h!] 
\centering
\includegraphics[width=0.45\textwidth]{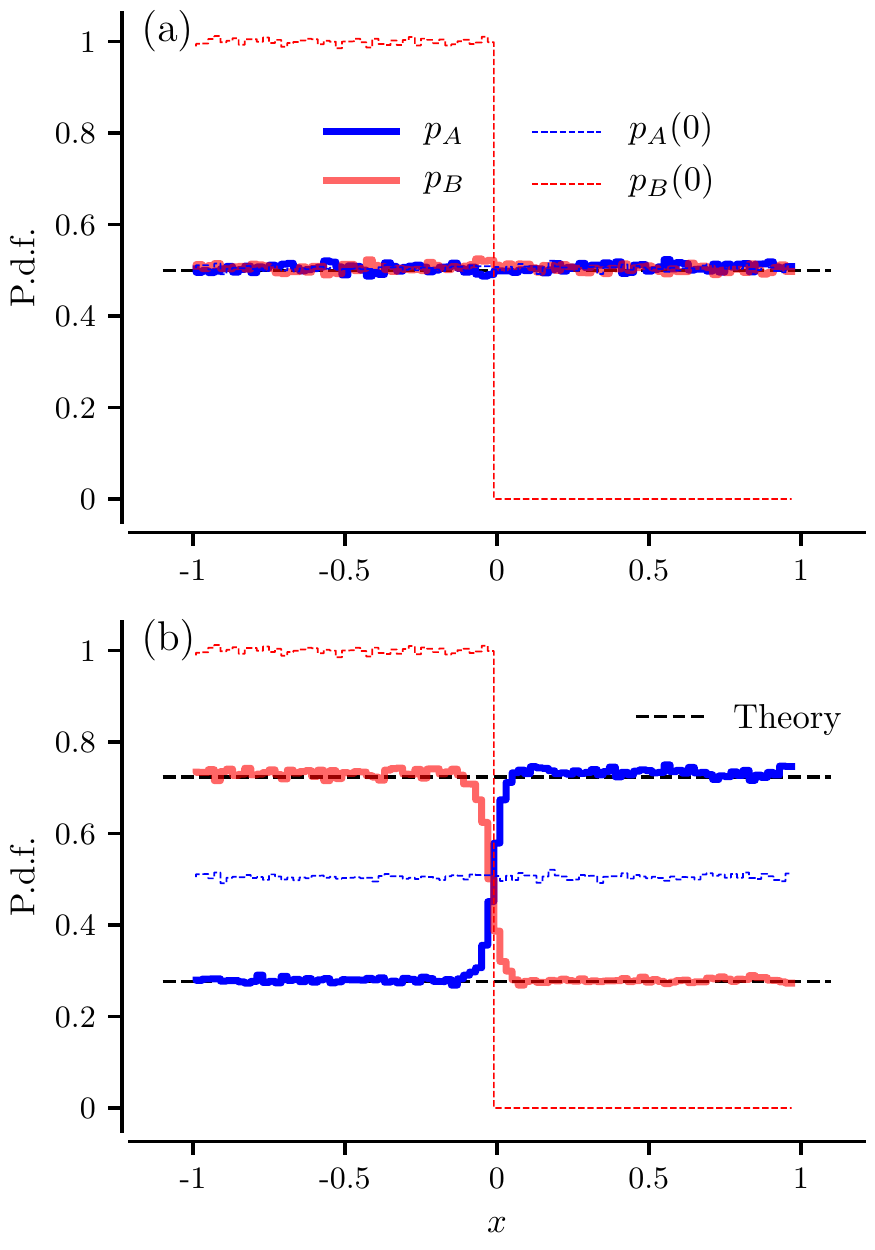}
\caption{Demixing of two species in simulations of Langevin equations \e{langevin_H}. Snapshots of the steady-state probability density functions (p.d.f), $p_{A,B}(x,t=10)$, are shown by solid lines for a subcritical value $c=3<c_\textrm{crit}$ (a) and a supercritical value $c=5>c_\textrm{crit}$ (b). Bin size for pdf estimation $\Delta x= 0.02$. Thin lines show corresponding initial densities $p_{A,B}(x,t=0)$: homogeneous and inhomogeneous (with an excess on the left) for $A$ and $B$ particles, respectively, and identical in (a) and (b). Theory, \e{sol_symm}, is shown by black dashed lines. Other parameters: number of particles $N=10^6$, integration time step $\Delta t=10^{-5}, s=1$.}
	\label{fig:simu_nonlinear1}
\end{figure}

In \bi{simu_nonlinear1} we demonstrate the existence of the demixed stationary state of the two species for a supercritical control parameter $c$. Starting with a step-like inhomogeneous  distribution of $B$ particles (thin red line) and a homogeneous distribution of $A$ particles (thin blue line), we simulate the system until the steady-state pdfs do not appreciably change anymore. For a subcritical nonlinearity ($c<c_{\rm crit}=4$) as in \bi{simu_nonlinear1}a, the two densities (thick blue and red lines) both approach a uniform profile $p_A(x)\,=\,p_B(x)\,=\,1/2$  - no demixing is observed in this case. In contrast, the two species prepared in  the same initial state as before but for a stronger nonlinearity ($c>c_{\rm crit}=4$) settle in an inhomogeneous steady state (\bi{simu_nonlinear1}(b)), in which the excess of one particle sort is accompanied by a shortage of particles of the other sort.
The two pdf-profiles are symmetric with respect to $x=0$ where a sharp interface separates the two populations. The higher value of $p_B$ on the left is (approximately) equal to the higher level of $p_A$ on the right; the same holds true 
for the lower values. The deviations from a uniform density can be well predicted by the theory (dashed lines) that is detailed below in \se{symm}. 

\begin{figure}[h!] 
\centering
\includegraphics[width=0.45\textwidth]{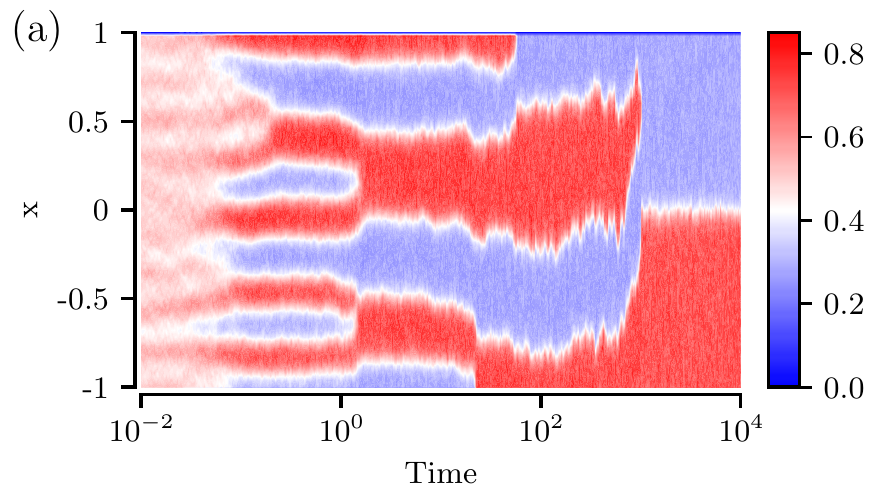}
\includegraphics[width=0.45\textwidth]{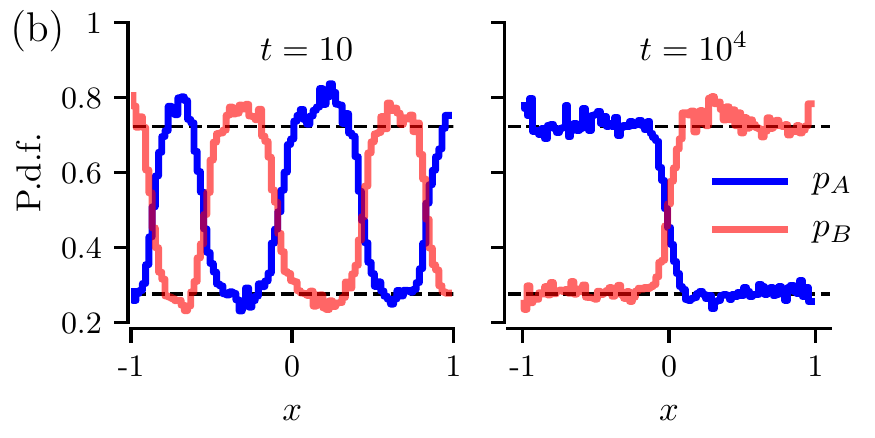}
\caption{Transient dynamics of demixing. Snapshots of probability density functions for $c=5$, $s=1$, bin size for pdf estimation $\Delta x= 0.02$, and uniform initial distributions of particles.	(a): Heat map of $p_A(x,t)$; values of the p.d.f. are according to the color-bar.	(b): Snapshots of probability density functions at $t=10$ and $t=10^4$ (final state of graph (a)). Black dashed line shows theoretical upper and lower bounds (\ref{eq:sol_symm}).	Other parameters are: $c=5$, number of particles $N=10^5$, integration time step $\Delta t=10^{-4}$.
	}
	\label{fig:simu_nonlinear}
\end{figure}

We note that with periodic boundary conditions (not shown) the configurations look similar but  a second interface is created. Whereas the interface for the used reflecting boundary condition is fixed on average to $x=0$, for periodic conditions, the interfaces can move stochastically due to the existence of the Goldstone mode (the distance between the interfaces is approximately constant).

Even if for reflecting boundary conditions, the position of the interface seems to be fixed, the steady-state solution is still strongly influenced by the initial conditions. Because the system is completely symmetric with respect to $A$ and $B$ particles, it is evident that an excess of $B$ particles on the right and an excess of $A$ particles on the left should be also a steady-state solution for the system. Our initial condition that started with an excess of $B$ particles on the left seems to promote the evolution towards a steady state in which $B$ particles are still in excess on the left.   What happens if we do not bias the system by the initial condition?

In turns out that not only \emph{which} solution but also how \emph{quickly} this steady solution is approached, depends strongly on the chosen initial distributions. In \bi{simu_nonlinear} we present simulations results in which \emph{both} densities were started in an spatially uniform state, $p_A(x.t=0)\,=\,p_B(x,t=0)\,=\,1/2$. Here we also show the time-dependent probability density for the $A$ particles (heat map in \bi{simu_nonlinear}a), illustrating that the uniform initial densities lead for short times to a large number of interfaces; both densities jump between the two (theoretically predicted values) back and forth, such that an excess of one sort of particles implies an a scarcity of the other sort.
As time goes on, the number of interfaces drops slowly - inhomogeneous domains will approach each other and merge. This, however, is a slow process and in our simulation it takes more than a span of  $t = 1000$  for the system to settle in the ultimate steady state exhibiting only one interface.
 We note that in contrast to the tendency of the Langevin system to minimize the number of interfaces, the local Smoluchowski density equations studied below in \se{mf} admit an arbitrary number of interfaces (jumps between discrete levels) in their steady state solution (cf. \bi{initcond}).

\begin{figure} 
\centering
\includegraphics[width=0.45\textwidth]{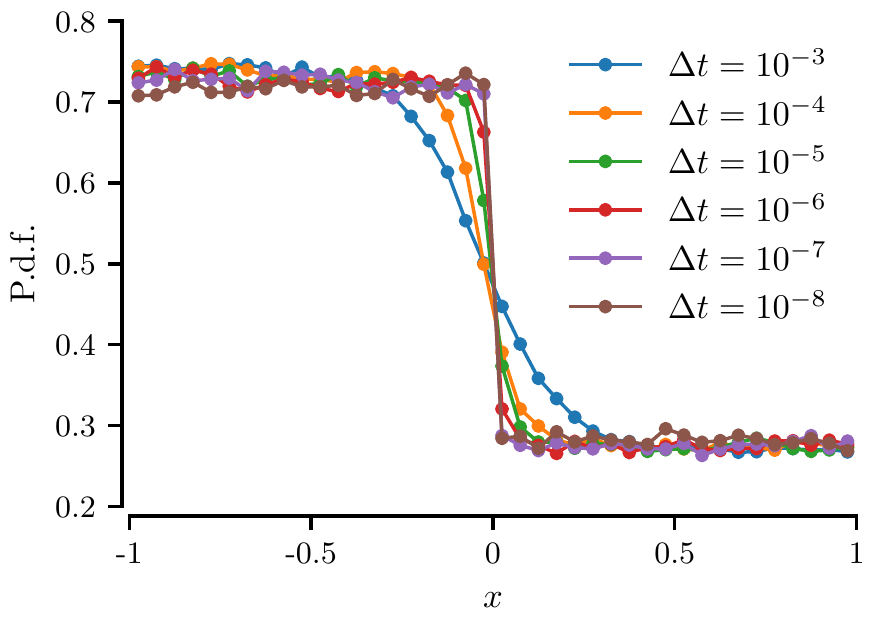}
\caption{Sharp density interface of demixing. Steady-state probability density functions for the indicated values of integration time step, $\Delta t$. Other parameters are:  $c=5$, number of particles $N=10^6$; $s=1$, bin size for pdf estimation $\Delta x=0.05$. Integration time $t=10$.}
\label{fig:simu_dt}
\end{figure}

Below, we will show results from numerical integrations of the corresponding mean-field Smoluchowski equations for the steady-state pdfs. In marked contrast to the particle simulations, it will turn out that these density equations can maintain a considerable number of jumps. With an extension of the sensing domain larger than a grid element $s>1$ we observe a coalescence process to a single interface. 

Turning back to the particle simulations we would like to point out that not only the number of interfaces but also the exact shape of the profile will depend on the details of the numerical procedure. This is illustrated in \bi{simu_dt} where  we investigate the sharpness of the interface depending on the used time step of our integration scheme. Remarkably, the interface can become extremely sharp: starting with time steps about $\Delta t \approx 10^{-5}$ and smaller, the density profiles exhibit macroscopic jumps between the adjacent grid elements around $x=0$. Only for time steps $\Delta t<10^{-6}$, the density anticipates the piecewise constant function that we will find in the next section as the solution of a partial differential equations (cf. below \e{solution} and \bi{initcond}).

\section{Mean field equations with arbitrary sensing radius}	
\label{sec:mf}
In the macroscopic limit $N\to \infty$ and $\Delta x \to 0$ (while keeping $s \Delta x$, i.e. the sensing radius, constant), the set of Langevin equations corresponds to a Fokker-Planck equation for the pdf of the $2N$ particles of our ensemble. The latter is a high-dimensional diffusion equation in the $2 N$-dimensional position space. Since the diffusion coefficients depend on the current locations of the particles via the $\mean{\Pi_{A,B}}^q_{r_s}$ we have to state how to interpret the stochastic differential equation. Because the considered particles are active objects, their intrinsic noise arises from variations of their internal propulsion mechanism. We assume here that this mechanism contributes in a temporally discrete fashion very different to the thermal noise acting on passive Brownian particles in fluids. Consequently,
we have to use the Ito rule \cite{hanggi1982stochastic,risken1996fokker} for the formulation of the kinetic equation of the pdf (the Stratonovich calculus for limits of smooth increments is discussed in the appendix \se{eigenvalues_and_strato}).

Starting with the full probability density $P=P_{2N}(\ldots,x_{A,i},
\ldots,x_{B,j},\ldots;t)$ for the $2N$ particles, we obtain in Ito interpretation:
\begin{eqnarray}
&&\frac{\partial P}{\partial t}  \,=\,\sum_{i=1}^{N} \, \frac{\partial^2}{\partial x^2_{A,i}}\bracket{1\,+\,c \,\mean{\Pi_B(x_{A,i})}^q_{r_s}}\,P\\
&& ~~~~~~~~~+\,\sum_{j=1}^{N} \, \frac{\partial^2}{\partial x^2_{B,j}}\bracket{1\,+\,c\,\mean{\Pi_A(x_{B,i})}^q_{r_s}}\,P\nonumber
\label{eq:ito_smol}
\end{eqnarray}
Reduction to the one-particle pdfs $p_A(x;t), p_B(x;t)$ of an arbitrary $A$ and $B$-particle at position $x$ is performed by integrating over all possible values of the other positions, by de-correlating the particles in a mean field approximation and by using \e{link_pdf}. This yields the nonlinear and nonlocal set of coupled Smoluchowski-equations
\begin{eqnarray}
&&\partial_t\,p_A(x;t)\,=\,\partial_x^2\,\bracket{\,1\,+c \,\mean{p_B}^q_{r_s}}\,p_A,\nonumber\\
&&\partial_t\,p_B(x;t)\,=\,\partial^2_x\,\bracket{\,1\,+\,c \,\mean{p_A}^q_{r_s}}\,p_B\,.
\label{eq:smolu}
\end{eqnarray}
These are the basic equations for the further numerical and analytical exploration of the system. The boundaries are as in the particle simulations, either reflecting or periodic at $x\,=\,\pm \ell$. Complications may arise in case of nonvanishing sensing radii which have to be taken into account in the formulation of the boundary conditions.
In particular, when solving \e{smolu} on a discrete grid, one has to extend the number of grid elements beyond the boundaries corresponding to one sensing radius.  

As \e{smolu} are nonlinear and nonlocal, multiple stationary solutions might exist. The stability of these solutions may change when changing parameters (bifurcations of the steady solutions). The simplest guess for stationary solutions are  two uniform distributions $p^0_A=p^0_B=p_0=1/(2\ell)$, i.e. a situation in which both types of particles are well mixed. In the next subsection we will address the stability of this uniform state.

\subsection{Stability analysis}
\label{sec:stability}
We modify the stability analysis of one-component systems in Refs.~\cite{lopez2004fluctuations,hernandez2004clustering,lopez2005self} in order to make it applicable to the case of two species. We assume small periodic perturbations $\delta p_{A,B} \propto \exp(\lambda t +{\rm i} kx)$ around $p_0$ with eigenvalue $\lambda$ and wave number $k$. Corresponding to the normalization condition for the probability densities and the boundary conditions, we choose $k\,= \,l\, \pi, ~~l\,=\,1,2,\ldots$ and $x\in [-1,1]$. Linearisation of the density equations with respect to the small perturbations  yields the dispersion relation for the larger of the two eigenvalues
\begin{equation}
\lambda(k)\,=\, -k^2\bracket{1\,+\,c \,p_0^q \bracket{1 \,-\, q\frac{\sin(kr_s)}{k r_s}}}\,.
\label{eq:disp_gen}
\end{equation}
Let us first consider the simple limit of a vanishing sensing radius, in the continuous case of  \e{smolu} given by $r_s \to 0$; for this local case we give also the details of the stability analysis in the appendix \se{eigenvalues_and_strato}. We then obtain
\begin{equation}
\lambda\,=\,-\,k^2\,\bracket{1\,+\,c\,p_0^q (1-q)}\,.
\label{eq:disp_local}
\end{equation} 
Clearly, if $q=1$ the eigenvalue is always negative, i.e. there is no instability in this case. With $q=2$  and $p_0=1/2$ all eigenvalues become positive for coupling values larger than the critical value 
\be
c_{\text{crit}}=\frac1{p_0^q (q-1)}=4;
\label{eq:crit}
\ee
the latter value is obtained for our parameter choices of $\ell=1, q=2$. For values of $c$ above this critical value, the homogeneous solutions are destabilized and the fastest growing mode at the critical situation is $k \to \infty$. Hence, when perturbations of all wave lengths occur, we will first see an instability corresponding to very high wave number or correspondingly very small wave lengths, i.e. the homogeneous mixture  of the two sorts of particles decomposes starting with tiny spatial separating regions. 
\begin{figure}[t!]
	\centering 
	\includegraphics[width=0.5\textwidth]{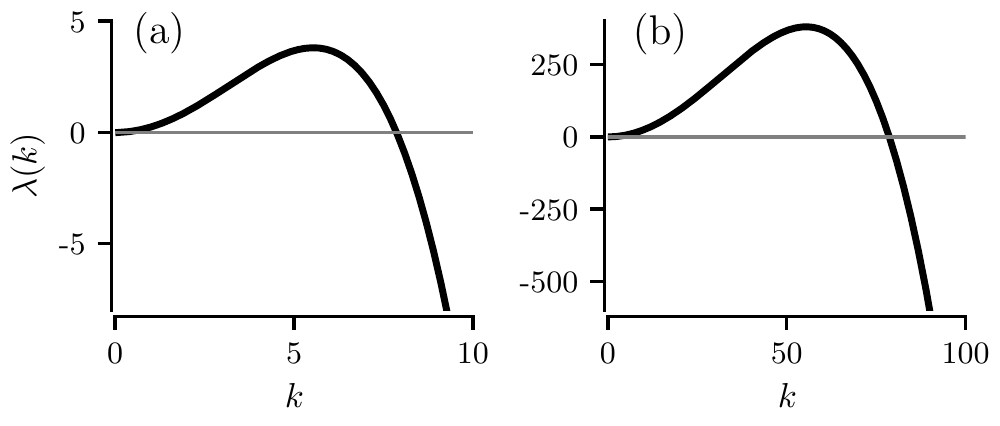}
	\caption{Eigenvalue vs $k$ according to \e{disp_gen} for $c=5$ and different values of the sensing radius: (a) $r_s=0.1$; (b): $r_s=0.01$. Note that only discrete values of $k=l \pi$ with $l=1,2,\cdots$ can be attained.}
    \label{fig:eigenvalue}
\end{figure}

In contrast and as can be expected, a finite sensing radius $r_s>0$ evokes a spatially more smooth destabilization of the homogeneous solution according to \e{disp_gen}.
First of all, it can be shown that $\lambda(k;r_s>0)<\lambda(k;r_s\to 0)$; secondly, the difference between vanishing and non-vanishing sensing radius grows with $k$ and 
$\lambda(k;r_s>0)$ will become negative for sufficiently large $k$ even for $c>c_{\text{crit}}$. This is illustrated in \bi{eigenvalue} where we show the eigenvalue $\lambda(k)$ vs $k$ for two different sensing radii as indicated in the caption (note that despite the continuous curve only values at $k=\ell \pi$ have to be considered). For smaller value of $r_s$ (\bi{eigenvalue}b) the fastest growing modes are found at larger wave number $k$ in line with results of the local theory where the fastest mode is at $k\to \infty$. Also the growth velocity of perturbations becomes larger (cf. scales of the $\lambda$ axes in \bi{eigenvalue}a and b).

There is a second transition for fixed $c$ with growing sensing radius $r_s$. Let $c>c_{\rm crit}$. If $r_s$ becomes comparable to the overall length scale $2\ell$  of the considered situation, the uniform distribution resumes stability. The eigenvalue again changes the sign and for large sensing radii the single stable solution is the uniform one. Taking the lowest possible value of $k=\pi$, we obtain for the critical radius the equation
\begin{equation}
1+c p_0^q (1- q\frac{\sin(\pi r_s)}{\pi r_s})=0 \; \rightarrow \; \sin(\pi r_s)= \frac{(c p_0^q)^{-1}+1}{q} \pi r_s
\label{eq:stab_uni}
\end{equation}
The solution will obviously depend on the value of $c$ but we can ask what happens if we have an arbitrary strong coupling ($c\to \infty$). In this case, the solution of the transcendental equation for $q=2$ is $r_{s,\text{crit}}\approx 0.6034$, i.e. for values of the sensing region larger than 61 \% of the system size we can exclude any instability.

We found that for $c>c_\text{crit}$ and sufficiently small sensing radius,  homogeneous densities of each particle sort become unstable, however, does that also imply an inhomogeneity for the \emph{total} distribution of particles?  
This question can be addressed by inspecting the stability of the overall density $p(x,t)=p_A(x,t)+p_B(x,t)$. The latter is normalized to $2$ and the steady uniform distribution reads $p(x)=1$. We can write down equations for $p(x,t)$ and for the density of one sort of  particles, say $p_B(x,t)$:
\begin{eqnarray}
\partial_t p(x,t)&=&\partial_x^2 [p\,+\,c\,((p-p_B)\,\mean{p_B}^q+p_B\,\mean{p-p_B}^q)]\,,\nonumber\\
\partial_t p_B(x,t)&=&\partial_x^2 [(1\,+\,c\,\mean{p-p_B}^q)\, p_B(x,t)]\,.
\end{eqnarray}
and everywhere $p_A=p-p_B$ have to be inserted. The eigenvalues of the equations linearized with small $\delta p$ and $\delta p_B$ around the steady state solutions $p=1,p_B=p_0=1/2$ factorize which results for the eigenvalues of the overall density $p$
\begin{equation}
\lambda_p(k)=\, - k^2 \bracket{1+\,c \,p_0^q\,\bracket{1\,+\,q \frac{\sin(k r_s)}{k r_s}}}\,
\label{eq:eigen_over}
\end{equation}
and the expression known from \e{disp_gen} the pdf of $B$-particles. For $q=2$, the perturbations of $p(x,t)$ decay with eigenvalue $\lambda_p$ (the factor $1+2\sin(k r_s)/(k r_s)$ remains strictly positive) and the uniform distribution of the sum of the pdfs with $p=1$  is always stable; this will be different for stronger nonlinearities $q>2$ but is not explored here any further. Below, we will make use  of the overall density's asymptotic stability  around the uniform steady state.

\subsection{Symmetric asymptotic cases: Numerical findings}
\label{sec:numerics0}
We consider the asymptotic steady states  obtained from the long-time limit of the numerical integration of the local coupled Smoluchowski-equations; the nonlocal model will be considered in \se{nonlocal}. The local model also allows for an analytical solution that we explore in the subsequent subsection.
 
\begin{figure}
	\centering 
	\includegraphics[width=0.45\textwidth]{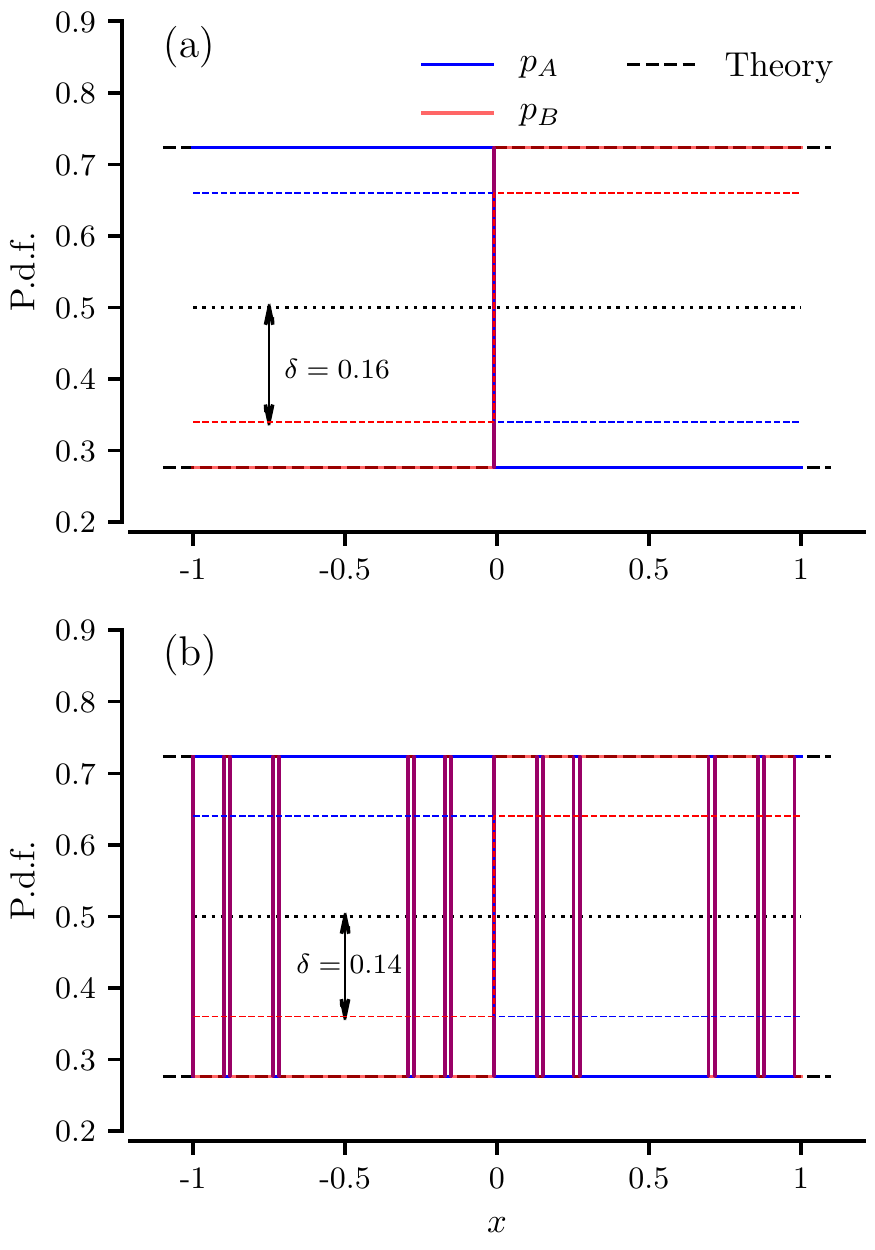}
	\caption{Symmetric stationary solutions from numerical solutions of the local Smoluchowski equation and from analytical treatment with reflecting boundary condition. In both panels the initial and stationary solutions are shown by dashed and solid colored lines, respectively. Black dashed lines show theoretical predictions \e{sol_symm} (see next subsection). Panels differ in their initial distance to the uniform distribution (dotted line):  $\delta=0.16$ (a),  $\delta=0.14$ (b). Other parameters: $M=100$ (number of grid points), $c=5$.}
	\label{fig:initcond}
\end{figure}

We start with the numerical integration results for the local case, $r_s=0$.  The parameter $c$ is adjusted to values where the uniform distribution is unstable. In particular we use $c=5$ if not stated otherwise. We use initial conditions for $p_{A,B}(x,0)$ in form of a  step-function, which is symmetrical around the uniform state and around the origin at $x=0$, 
\begin{eqnarray}
&&p_A(x,0) = \displaystyle 
\begin{cases}
\frac{1}{2}+\delta,  \mbox{$-1\le x \le 0$}\\
\frac{1}{2}-\delta, \mbox{$0< x \le 1$}
\end{cases} \nonumber\\
&&p_B(x,0)=1-p_A(x,0). 
\label{eq:initcond}
\end{eqnarray}
The parameter $0 \le \delta \le 0.5$ specifies how far from the uniform state the initial conditions are. 

In \bi{initcond} we show representative numerical examples which exhibit demixed states. In \bi{initcond}a  the  particle species $A$ ($B$) displays an increased probability density  with numeric value about $p_h \approx 0.72$ in the domain left (right) from the origin. In contrast, right (left) from the origin the densities attain a diminished value $p_l \approx 0.28$.

In this case the initial conditions (colored dashed lines) differ sufficiently strongly from the uniform distribution and also trigger with their left/right asymmetry the asymmetry of the asymptotic solution. For the latter, deviations from 
uniformity, i.e. the values $p_h$ and $p_l$ of increased and diminished probability attained in the long-time limit (black dashed line) can be well predicted by the calculation presented in the next subsection. Note that the domains of increased and diminished probability have equal size and that an increase in one sort's pdf is accompanied by the decrease in the other sort's pdf (demixing). In line with the stability of the homogeneous state discussed in the previous section, we observe numerically that for arbitrary $x$
\begin{equation}
p(x)=p_A(x)+p_B(x)\,=\,p_h\,+\,p_l\,=\, 1\,.
\label{eq:overall}
\end{equation} 
 \bi{initcond}b illustrates  the drastic change caused by initializing the system closer to the uniform density (cf. colored dashed lines) by choosing a slightly  smaller value of $\delta$. Asymptotically, the same constant values $p_h$ and $p_l$ will be attained. However, in contrast to \bi{initcond}a, the two densities jump multiple times between these levels obeying strictly the demixing property that an increase in one sort comes along with a decrease in the other one. How many of such jumps can we observe and how does their number depend on the initial conditions for the two densities?

\begin{figure}[t!]
	\centering 
	\includegraphics[width=0.5\textwidth]{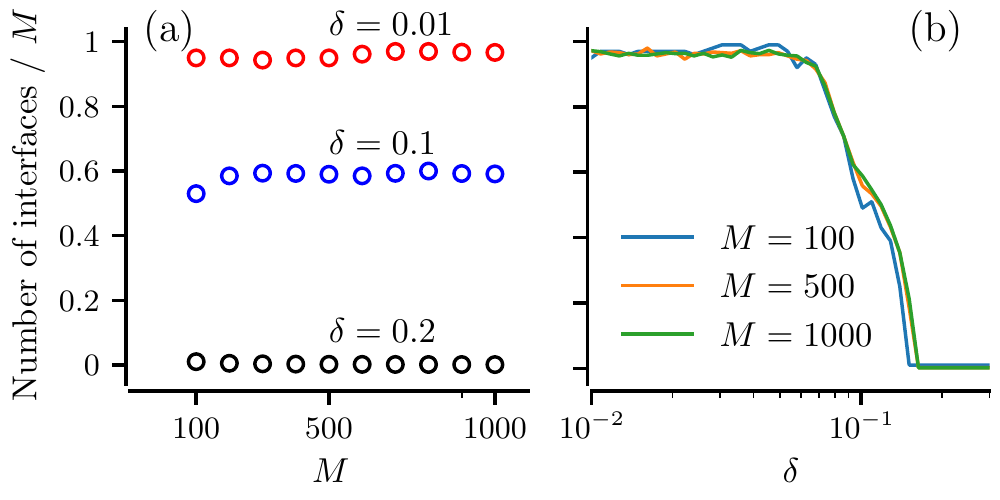}
	\caption{Total number of interfaces in the stationary pdfs vs the grid size and the initial condition parameter $\delta$.
	(a): The number of interfaces normalized to the grid size $M$ vs $M$ for the indicated values of $\delta$.
	(b): The normalized number of interfaces vs $\delta$ for the indicated values of $M$. 
    For both panels, $c=5$; the initial conditions are given by \e{initcond}.
    }	
    \label{fig:init-del}
\end{figure}

In the numerical solution of the coupled Smoluchowski equations we use finite grid elements and thus the maximal number of interfaces cannot exceed this number of elements. Any number of jumps below this maximal number  is possible in the local version of the problem (as long as the densities also  obey the normalization condition). This is  due to the absence of a surface tension, no activation is needed to create a couple of new interfaces (jumps). All these configurations with different numbers of interfaces appear asymptotically stable for the set of deterministic Smoluchowski \e{smolu}. 

Importantly, the precise number and localization of interfaces is as solution of the Smoluchowski equation uniquely determined by the given initial condition. The number of interfaces in the stationary pdfs depends on how far away from the uniform distributions the initial pdfs are, i.e. on parameter $\delta$ in \e{initcond}. As illustrated in \bi{init-del}b, the number of interfaces drops  monotonically  from $M$ (small $\delta$), the number of grid points,  to $1$ (for large values of $\delta$). 

The fraction of possible interfaces does not seem to depend much on the grid size once a sufficient size has been reached:  as demonstrated in \bi{init-del} the relative number of jumps (compared to the maximally possible number) depends on the initial distance to the uniform distribution, $\delta$ but not much on the number of grid points used. In contrast, as can be also extracted from Fig. \ref{fig:init-del}, the absolute number of jumps does depend on the number of grid elements. A different partition (i.e. a different $\Delta x$) of the same line $[-\ell,+\ell]$ creates another spatial arrangement of A and B. It possesses  the same density plateaus but usually a different number of interfaces.

%%%%%%%%%%%%%%%%%%%%%%%%%%%%%%%

Interestingly, only the solution with a single interface is in agreement with the long-time asymptotic solutions in particle simulations; we recall that solutions with multiple jumps  are observed as long-living transients (cf. \bi{simu_nonlinear}). Similarly to what we have observed there, an initial condition closer to a uniform distribution typically enables a larger number of interfaces.

\subsection{Symmetric asymptotic cases: Analytic findings}
\label{sec:symm}
Setting the temporal derivative to zero, we obtain  the stationary Smoluchowski equations in the local case
\begin{eqnarray}
&&\frac{d^2}{d x^2}\,(1\,+\,c\, p^2_A(x))\,p_B(x)\,=\,0\,, \\
&&\frac{d^2}{d x^2}\,(1\,+\,c\,p^2_B(x))\,p_A(x)\,=\,0\,.
\label{eq:stat_smolu}
\end{eqnarray}
Focussing on the case with reflecting boundary conditions, we can safely assume vanishing probability currents for the two species, and hence not only the second derivative but already the first derivative of the product has to be zero.   

Both densities must be thus (piecewise) constant, attaining a high level $p_h$ and a low level $p_l$, respectively. We assume in this subsection that the corresponding integration  constant in the two Smoluchowski equations is the same: 
\begin{equation}
 (1+c p^2_h)\,p_l \,= \,E\,=(\,1+cp^2_l)\,p_h\,.
\label{eq:stat_smol}
\end{equation} 
The assumed symmetry also implies that the domains of increased and decreased density are of equal size. The above equations together with the normalization condition provide algebraic conditions for the determination of the constant values $p_h$ and $p_l$ as solutions  left and right from interfaces (jump points).  We underline, due to the symmetry of \e{stat_smolu}, both $(p_l,p_h)$ and $(p_h,p_l)$ are  possible solutions for $(p_A(x), p_B(x))$ (and both have to be attained at least in two distinct domains to satisfy the normalization condition).

In order to find  analytical expression for $p_l,p_h$ we divide \e{stat_smol} by the product $p_h \, p_l$ and obtain
\begin{equation}
\frac{1}{p_h}\,+\, c\,p_h\,=\,\frac{1}{p_l}\,+\,c \,p_l
\label{eq:fact}
\end{equation}
Furthermore we can use that in agreement with \e{overall}, the two values should deviate by the same amount from the uniform density, $p_{h,l}=1/2\pm\epsilon$. One obtains quickly the first, trivial solution $\epsilon \,=\,0$ (both densities equal to the uniform distribution). The other solutions read 
\begin{equation} 
p_{h,l}\,=\,\frac{1}{2} \,\pm\,\sqrt{\frac{1}{4}\,-\-\frac{1}{c}}\,.
\label{eq:sol_symm}
\end{equation}
In line with the stability analysis of \se{stability} these solutions exist for sufficiently large $c$ beyond a pitchfork bifurcation at $c_{\rm crit}=4$. 

The simplest solutions  for reflecting boundary conditions and a supercritical value of $c$, i.e. the solutions with only one jump, can be formulated in terms of the Heaviside function 
\begin{eqnarray}
\label{eq:solution} 
&& p_A(x)=p_h \theta(x) + p_l \theta(-x), \nonumber\\
&& p_B(x)=p_l \theta(x) + p_h \theta(-x).
\end{eqnarray}
Of course, there is the second solution
\begin{eqnarray}
\label{eq:solution2} 
&& p_A(x)=p_l \theta(x) + p_h \theta(-x), \nonumber\\
&& p_B(x)=p_h \theta(x) + p_l \theta(-x).
\end{eqnarray}
in which both densities switch roles.

In principle,  normalized  densities with any (even or odd) number of jumps between $p_h$ and  $p_l$  are possible as long as the total domain size for one specific level (say, $p_h$) adds up to $\ell=1$. To give an example, the following solution would also satisfy the stationary Smoluchowski equations:
\begin{eqnarray}
\label{eq:solution3} 
 p_A(x)&=& p_l \theta(x_2-x_1-x)+p_h \theta(x+x_1-x_2)\theta(x_1-x)\nonumber\\
&& +p_l \theta(x-x_1) \theta(x_2-x)+p_h \theta(x-x_2), \nonumber\\
 p_B(x)&=&p_h \theta(x_2-x_1-x)+p_l \theta(x+x_1-x_2)\theta(x_1-x)\nonumber\\
&& +p_h \theta(x-x_1) \theta(x_2-x)+p_l \theta(x-x_2),
\end{eqnarray}
with $0<x_2<1$ and $-1<x_1<x_2$ and there can be two or three jumps of the density, depending on the choice of $x_1$ and $x_2$. However, like the numerical solutions our analytical solution allows the maximal number of jumps $M$ if the resulting function is in agreement with the normalization condition. The densities jump in antiphase between the two values $p_h$ and $p_l$  as calculated below. Every grid element $\Delta x$ is bound to attain one of two values.  Also configurations with extended pieces of constant densities between two interfaces are in agreement with the analysis. Note that for the solutions discussed so far, the number of grid elements with diminished density equals the corresponding number of increased density  due to the normalization condition.

%%%%%%%%%%%%%%%%%%%%%%%%%%%%%%%%%%%%asymmetric steady states

\subsection{Asymmetric asymptotic states: Numerical and analytical findings}
\label{sec:asymm}
A different class of stationary solutions emerges for the initial conditions $p_A(x,0), p_B(x,0)$, which are not symmetric, in particular when \e{overall} is initially not fulfilled and $p_A(x,0)+ p_B(x,0)\neq 1$ (at least for some range of $x$).  Consider, for instance, the initial condition for $p_A(x,t)$ as above with parameter $\delta$, but set a uniform initial pdf for species B, $p_B(x,0)=1/2$. 
Exemplary stationary pdfs obtained from numerical solutions of the Smoluchowski equations are displayed in  \bi{asym} a,b. In \bi{asym} c  we show the common distribution $p(x)=P-A(x)+p_b(x)$.  for two different values of $\delta$. Both densities again approach piecewise constant solutions but now differ in the attained constant levels: in total, we have now two different pairs of solutions $(p_{A,l},p_{B,h})$ and $(p_{A,h},p_{B,l})$. We recall that we had before in the symmetric solution $p_{A,l}=p_{B,l}$ and $p_{A,h}=p_{B,h}$. Now, however, these values are not the same anymore, $p_{A,l}\neq p_{B,l}, p_{A,h}\neq p_{B,h}$. In addition, the domain size for the two pairs of solutions is not equal and it seems to depend on the pair of values attained  and on the initial condition.
 \begin{figure}[h!]
	\centering 
	\includegraphics[width=0.45\textwidth]{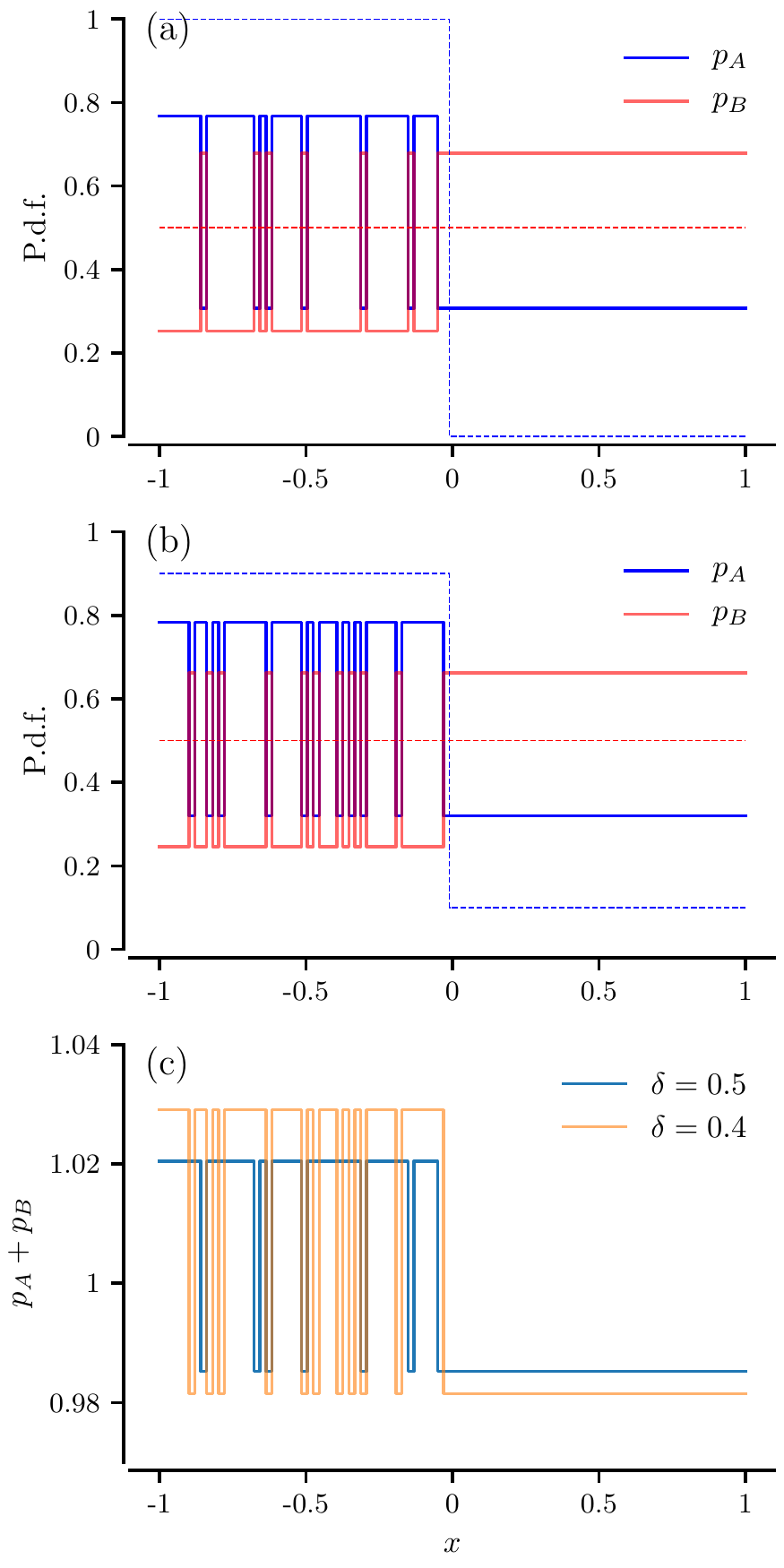}
	\caption{Asymmetric stationary solutions of the local Smoluchowski equation with different initial conditions in terms of parameter   $\delta=0.5$ (a) and  $\delta=0.4$ (b). Stationary solutions and initial conditions in (a,b) are shown by solid and dashed colored lines, respectively. We observe two pairs of piecewise constant solutions, $(p_{A,l},p_{B,h})$ and $(p_{A,h},p_{B,l})$ that occupy domains of distinct sizes $1+\Delta$ and $1-\Delta$, respectively; numerically, we find $\Delta=0.16$ (a) and $\Delta=0.22$ (b). Remarkably, these asymmetric solutions (asymmetric with respect to the domain sizes), lead to a slightly inhomogeneous distribution of particles in the two domains as shown in (c). Parameters: $M=100$ (grid points), $c=5$.}
	\label{fig:asym}
\end{figure}
  
In order to calculate these asymmetric solutions, we relax the assumption of equally sized domains of increased and decreased probability. With the new spatial asymmetry parameter $\Delta$, we have as new normalization conditions
\begin{equation}
p_{A,h}(1-\Delta)+p_{A,l}(1+\Delta)=p_{B,l}(1-\Delta)+p_{B,h}(1+\Delta)=1\,.\label{eq:normi_as}
\end{equation}     

\begin{figure}[h!] 
\centering
\includegraphics[width=0.45\textwidth]{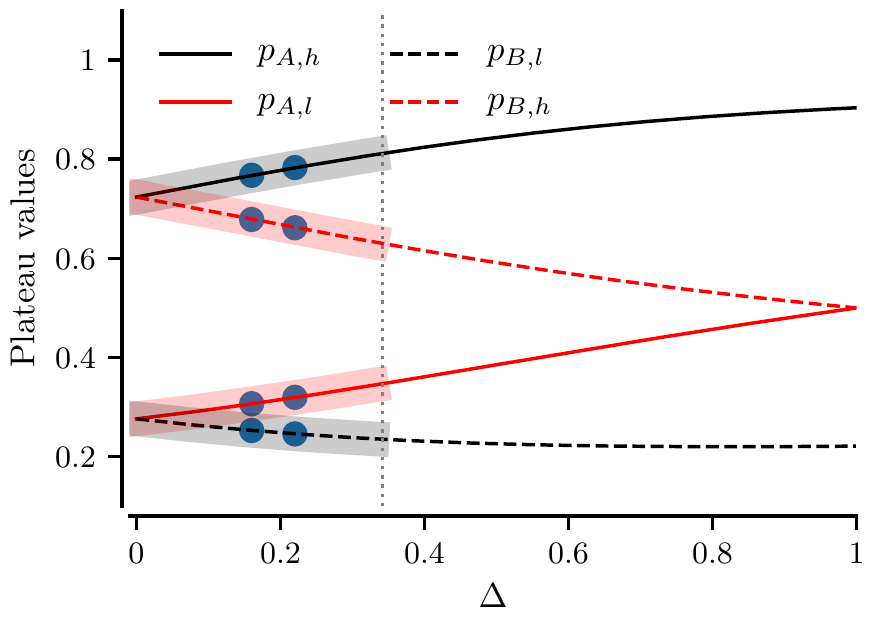}
\caption{
Stationary plateau values in case of asymmetric steps as
  function of $\Delta$. Numerical solutions of \e{stat_smolu_as} and the normalization condition \e{normi_as} for $c=5$. As revealed by the stability analysis in the appendix  \se{eigenvalues_and_strato}, the steady-state solutions are stable only for $\Delta<\Delta_{crit}\approx 0.34$ (indicated by thick lines). Blue points correspond to the asymptotic solutions found to be at $\Delta\approx 0.16$ in \bi{asym}a and at $\Delta\approx 0.22$ in \bi{asym}b.
  \label{fig:asymm_map}}
\end{figure}
Again from the Smoluchowski \e{smolu}, the
expressions under the Laplacians should be constants in the stationary
state. However, in contrast to the symmetric case, these integration
constants now possess different values
\begin{eqnarray}
&&(1+cp^2_{B,l})p_{A,h}\,=\,(1+cp^2_{B,h})p_{A,l}\,=\,E_A\,,\nonumber\\
&&(1+cp^2_{A,h})p_{B,l}\,=\,(1+cp^2_{A,l})p_{B,h}\,=\,E_B\,.
\label{eq:stat_smolu_as}
\end{eqnarray}  
Given a specific value of $\Delta$, we can solve the four equations \e{normi_as} and \e{stat_smolu_as} numerically for the four unknown levels. Results of this computation  as functions of the parameter $\Delta$ are presented in\bi{asymm_map} for $c=5$. The values for $\Delta=0$ correspond to the symmetric case \e{overall} and solutions \e{sol_symm} with $c=5$. In the limit $\Delta\to 1$, the surviving solution (the one for which the two levels are closer to each other and which is occupied in the larger domain $1+\Delta$) has to be a uniform distribution and, consequently, the dashed lines converge to 1/2. Because the uniform solution was already shown to be unstable for $c=5$, it is evident that not all values of $\Delta$ will lead to a stable asymptotic solution. In the appendix \se{eigenvalues_and_strato} we demonstrate  that sufficiently small $\Delta<\Delta_{crit}\approx 0.34$, the asymmetric solutions are stable with respect to weak perturbations (this range is indicated in \bi{asymm_map} by thick lines). For $\Delta<\Delta_{crit}$, one eigenvalue is positive and, hence, the corresponding state is not a stable solution.

Finally, we mention an important difference between the numerical solutions of the Smoluchowski equations and the particle simulation results for the  Langevin equations. First of all, asymptotically stable asymmetric states could \emph{not} be found in the Langevin simulations \e{diff_lang}. Hence, these solutions are a special feature of the Smoluchowski equations and a consequence of neglecting finite size fluctuations in the mean-field theory.  Inclusion of fluctuations at the level of the density equations might change this disagreement between the two levels of description. Secondly, because the Smoluchowski equations are deterministic, the solution for the same initial conditions will always be the same. This is not so, of course, for the Langevin equations -- even for very large populations of particles, a tiny fluctuation   may introduce a break in symmetry in one way or the other. The asymptotically stable states may differ for different runs of the system.

 \subsection{Inspection of stability of asymptotic case by Cahn Hilliard formalism}
We apply now another theoretical approach to the stability problem of the inhomogeneous state. The deterministic set of  local Smoluchowski-equations  can be regarded as a Cahn-Hilliard equation for a conserved order parameter \cite{fife2000models,bergmann2018active,speck2014effective,stegemerten2020bifurcations}. We write the Smoluchowski equations for the densities as nonlinear flux equations
\begin{eqnarray}
\partial_t\, p_A(x,t)\,=\,-\partial_x \,{J_A\{p_A,p_B]\}}\,,\nonumber\\
\partial_t\, p_B(x,t)\,=\,-\partial_x \,{J_B\{p_A,p_B\}}\,.
\end{eqnarray}
On the right hand side, there are the components of the flux-vector $J_{A,B}$, which can be expressed as functional derivatives 
\begin{eqnarray}
J_A\,=\,-\, \partial_x \frac{\delta \Phi}{\delta p_A}\,,~~~
J_B\,=\,-\, \partial_x \frac{\delta \Phi}{\delta p_B}\,.
\end{eqnarray}
for which the potential reads
\begin{eqnarray}
\Phi\{p_A(x,t),p_B(x,t)\}\,=\,\frac{1}{2} \int_{-1}^1\,{\rm d} x\,\left(p_A^2\,+\,p_B^2\,+\, c\,p_A^2\,p_B^2\right)\,.\nonumber\\
\label{eq:2pot}
\end{eqnarray}
This potential is locally defined. Because we consider the local version of the model, which has 
a vanishingly small sensing radius, the potential does not contain any interaction term corresponding to a creation of surface tension. The lack of such a term
is the reason for the narrow interfaces \cite{magaletti2013sharp,lee2016sharp}. 

It is well known that an potential like \e{2pot} plays the role of a Lyapunov function \cite{wilczek2015modelling}. Hence, we can use $\Phi$ to discuss the stability properties of the steady states. Insertion of the uniform state yields
\begin{equation}
\Phi_{\rm unif}\,=\, \frac{1}{2}\,+\,\frac{c}{16}\,.
\label{eq:pot_hom}
\end{equation}
It is compared to the potential value of the inhomogeneous solution with in  antiphase distributed $p_{A,B}=1/2 \pm \sqrt{1/4-1/c}$, both extended over a length of $1$. Here the dependence for $c>4$ yields 
\begin{equation}
\Phi_{\rm inh}\,=1-\frac{1}{c}\,.~~~~ c \geq c_{\rm crit}
\label{eq:pot_inh}
\end{equation}
Both curves intersect at $c=4$. However, for a supracritical value $c> c_{\rm crit}$ the potential value \e{pot_inh} is smaller and hence corresponds to the stable solution. We note that the value of $\Phi_{\rm inh}$ does not depend on the number of jumps (interfaces) which is a consequence of the local character of the potential.

Further on, we inspected the asymmetric steady states by help of the above mentioned Lyapunov function $\Phi\{p_A(x),p_B(x)\}$ within the parameter region $c > c_{\rm crit}$. By insertion of the spatially asymmetric inhomogeneous states, one finds  numerical values higher than these from the corresponding symmetric ones $\Phi\{p_A(x)_{\rm asym},p_B(x)_{\rm asym}\}>\Phi\{p_A(x)_{\rm sym},p_B(x)_{\rm sym}\}$.

Hence, we expect that the asymmetric configurations correspond to metastable states that are left quickly towards the symmetric states, once fluctuations are taken into account. This is exactly the case in the particle simulations and may explain why asymmetric states are not observed in the latter.

\section{Nonlocal model with finite sensing radius}
\label{sec:nonlocal}
Despite some agreement between the results of particle simulations and  of the corresponding mean-field theory in form of coupled Smoluchowski equations, we found also a number of striking discrepancies between these two levels of description.
One prominent difference, on which we focus now, is that in the Langevin simulations a state with several domains and corresponding interfaces develops asymptotically into a state with a single interface whereas the Smoluchowski equations with local coupling admit also asymptotically stable states with arbitrary number of interfaces, limited only by the number of grid points used in the numerical integration scheme. 
In addition, asymmetric states which do not obey \e{overall} also appear  in the deterministic (mean-field) treatment which we never observed in particle simulations.

\subsection{Transient and asymptotic states in the non-local model}

\begin{figure}[h!]
	\centering 
	\includegraphics[width=0.5\textwidth]{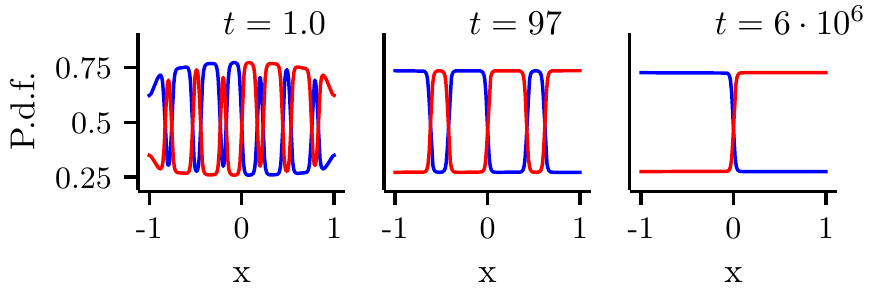}
	\caption{
		Evolution of the probability density functions towards stationary state for the non-local case for $c = 5$,  a spatial grid with $M= 500$, and a sensing radius, $r_s=0.01$. Both species are initially distributed according to step-functions  with a single interface shifted to negative values of $x$. Three panels show snapshots of $p_A(x,t)$ and $p_B(x,t)$ at indicated values of time $t$.}
	\label{fig:non-local_snapshots}
\end{figure}	

Integration of the Smoluchowski equation shows that the grid size $M$ determines the number of domains for the symmetric as well as for asymmetric distributions. This ambiguity of the numerical integration results from the local interaction in case of sensing radius $r_s=\Delta x/2$ as assumed above. As shown below, this ambiguity and the differences between particle simulations and mean field theory will be removed if the model includes a larger value of the sensing radius, specifically, equal or larger than the size of an individual grid element $r_s \ge \Delta x$.

A typical temporal evolution in case with  sensing radius $r_s\ge\Delta x$ is presented in \bi{non-local_snapshots}. Arbitrary initial configuration relax quickly to a symmetric state in which the full probability density $p(x)p_A(x)+p_B(x)$ is uniform, i.e. \e{overall} is obeyed, but still several interfaces coexist. The corresponding time scale  of this first relaxation of the initial state is given by the eigenvalue $\lambda_1$ from \e{eigen_over}. The distribution during this period resembles  the transient states with many interfaces observed in some of our particle simulations (cf. \bi{simu_nonlinear}) and the asymptotic solution for certain initial conditions obtained for the local Smoluchowski equations (cf. \bi{initcond}b). Eventually,   for very long times, the different domains coalesce and a single interface remains between two demixed states of $A$ and $B$ particles. This corresponds to the single inhomogeneous asymptotic state, which we found in particle simulations.

\begin{figure}
	\centering 
	\includegraphics[width=0.45\textwidth]{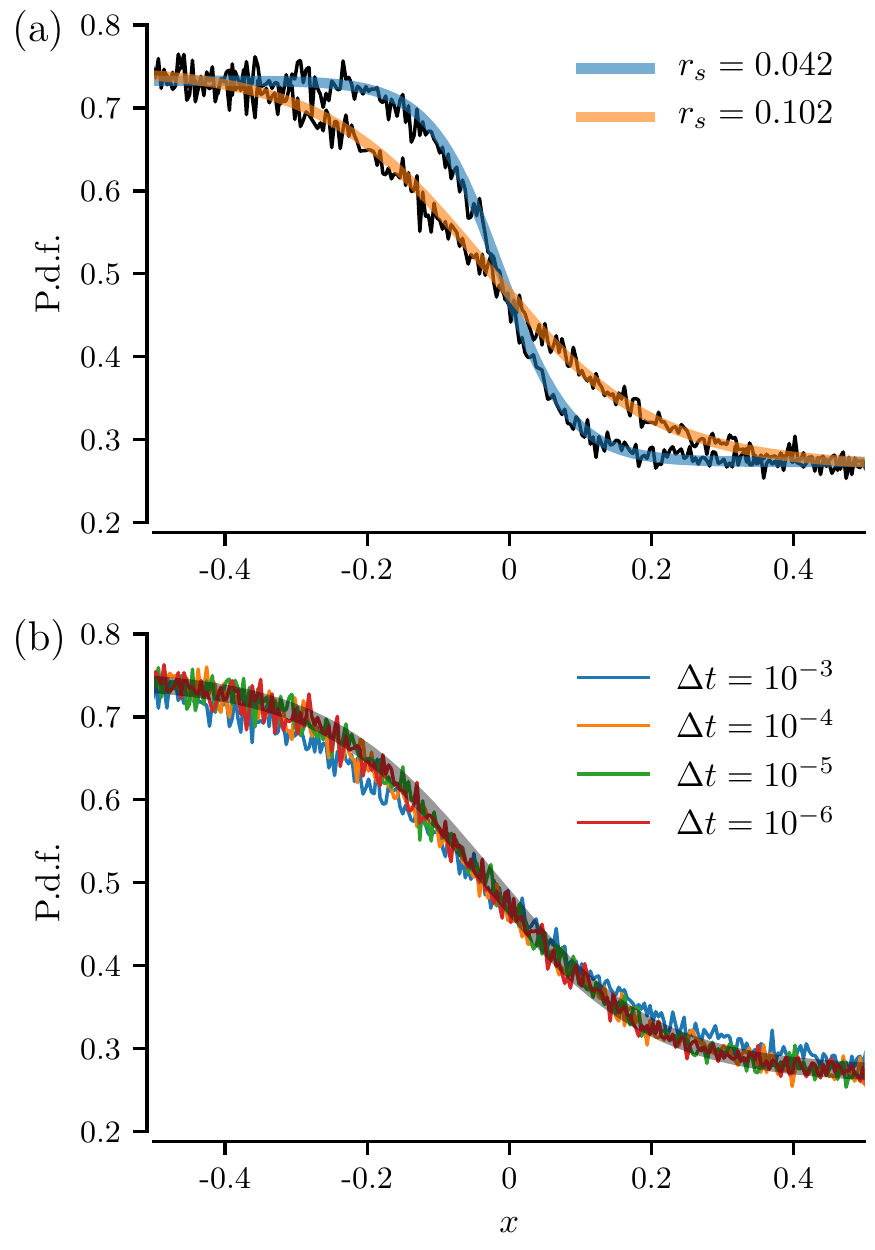}
	\caption{Stationary probability density, $p_A(x)$, for the non-local model estimated from Langevin equations, and from the numerical solution of Smoluchowski equation.	Initial conditions were given by \e{initcond} with $\delta=0.5$ for both Langevin and Smoluchowski equations. Number of particles of each species, $N=10^6$, and integration time $t=10$, for Langevin equation.  Number of grid points, $M=500$.
	(a): $p_A(x)$ from the Smoluchowski equation (thick color lines) and its estimate from Langevin equations (black lines) for the indicate values of the sensing radius, $r_s$.	The integration time step of Langevin equations, $\Delta t=10^{-5}$.
	(b): Comparison of the stationary solution of the Smoluchowski equation for $s=26$ (thick grey line) with estimates from Langevin equations for the indicated values of integration time step, $\Delta t$.
	} 
	\label{fig:nonlocal-frontprofile}
\end{figure}

Fig.~\ref{fig:nonlocal-frontprofile} illustrates the excellent correspondence between stochastic simulations of the Langevin equations (particle simulations) and  the numerical solution of the Smoluchowski equations. In particular, \bi{nonlocal-frontprofile}(b) demonstrates  convergence of particle simulations to the stationary solution of Smoluchowski equation already for relatively large time step, $\Delta t \approx 10^{-3}$. This can be contrasted with the local case of \bi{simu_dt}, where such convergence is observed for much smaller $\Delta t$.
The increase of sensing radius widens the profile of the smooth interface. Eventually, for the sensing parameter $s$ comparable with $M/2$ (i.e. $r_s=\ell=1$), the uniform solution becomes stable (not shown in \bi{nonlocal-frontprofile}(a)),  which we found already by inspecting the eigenvalue in  \e{stab_uni}. Upon further increasing  the  sensing radius, 
the governing Smoluchowski equations are less and less affected by the respective other species and thus loose their nonlinear character.

\begin{figure}[h!]
	\centering 
	\includegraphics[width=0.45\textwidth]{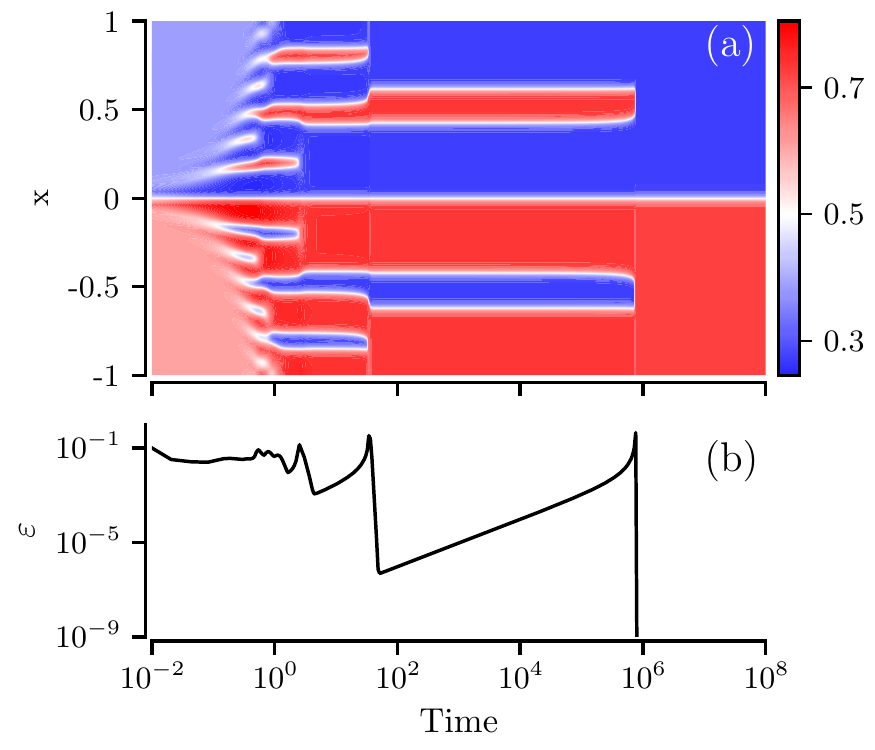}
	\caption{Evolution of the probability density functions from symmetric initial condition \e{initcond} with $\delta=0.12$, towards stationary state. Numerical integration used the spatial grid with $M=500$ and sensing radius, $r_s=0.01$. (a): The heat map of $p_A(x,t)$.
	(b): The convergence error, $\varepsilon(t)$.}
	\label{fig:non-local1}
\end{figure}

Generally, the long time needed to go to the final steady state depends crucially on the initial conditions.  When started from symmetric initial conditions  \e{initcond} with large $\delta$, solutions of both the Langevin and Smoluchowski equations approach stationary solutions quickly. In contrast, if the initial distribution \e{initcond} is close to the uniform, i.e. for small values of $\delta$, the transient to stationary solution may become extremely slow, as illustrated in the heat-map plot in \bi{non-local1}(a). As can be seen, multiple interfaces developed at small times converges eventually to the single interface, resulting in the stationary solution shown in \bi{nonlocal-frontprofile}b (blue line). This approach to the stationary solution can be quantified by the convergence error defined as the maximum of the difference of probability densities at two consecutive integration windows, $T$:
$$
\varepsilon (t) = \max \left| p_A(x,t) - p_A(x,t+T)\right|.
$$
Peaks in the convergence error shown in \bi{non-local1}(b) correspond to the merging of interfaces in the probability density.

\begin{figure}
	\centering 
	\includegraphics[width=0.45\textwidth]{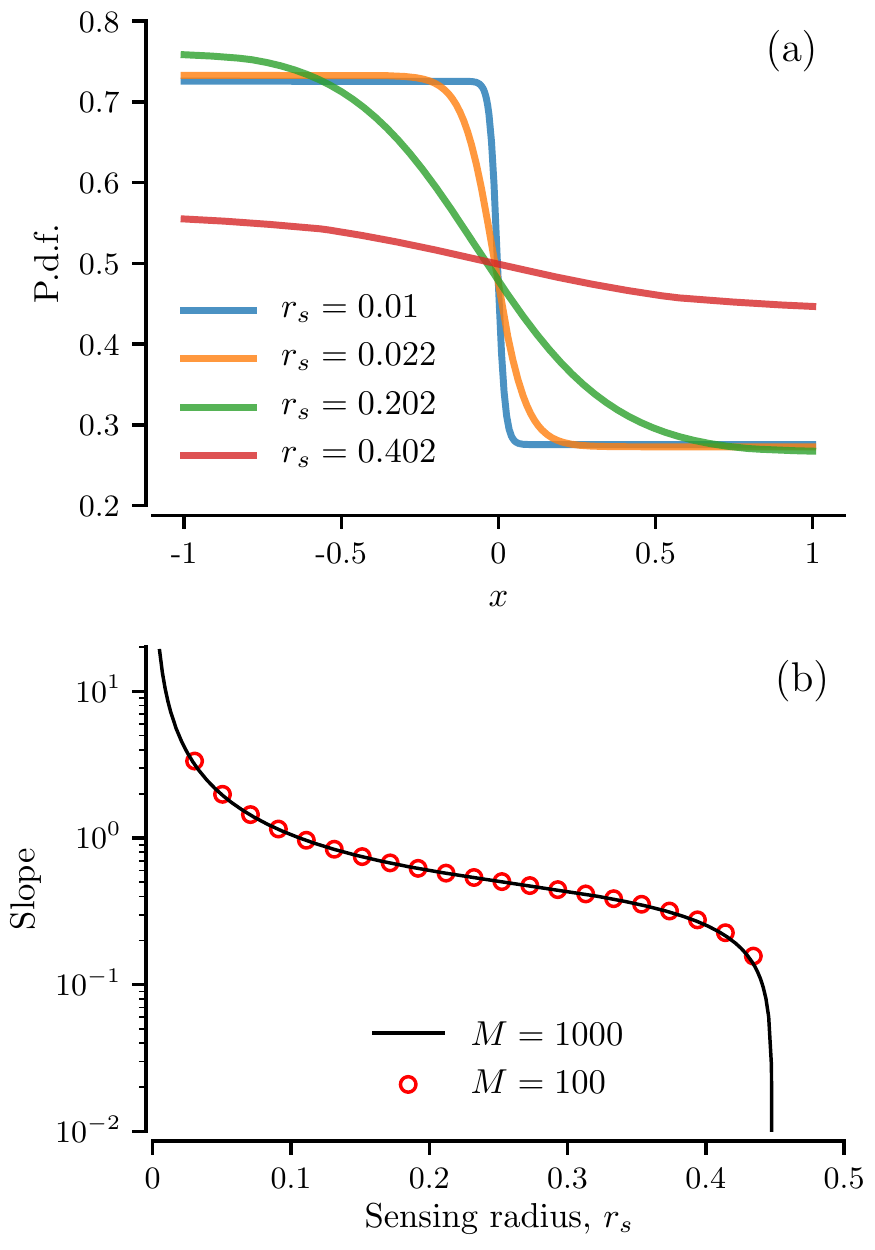}
	\caption{Effect of the sensing radius on the stationary probability density function obtained from numerical solution of the Smoluchowski equation.
		(a): Stationary p.d.f., $p_A(x)$, for the indicated values of sensing radius, $r_s$. Number of grid points, $M=500$.
		(b): Slope at the interface center  of the stationary p.d.f. vs sensing radius, $r_s$, for the indicated numbers of grid points, $M$. This figure indicates that for a given sensing radius the stationary solution is invariant with respect to the number of grid points, $M$. It changes from very steep occurrence for small radii to very flat interfaces for large nonlocal sensing.}
	\label{fig:nonlocal-summary}
\end{figure}

In \bi{nonlocal-summary} we summarize the properties of the asymptotic state if crossing from the local to the nonlocal model. First of all, already a size of a few grid elements $\Delta x$ are sufficient to remove  the strong dependence on the initial state with its multiple interfaces and the ambiguity of the local model, in which the asymptotic state is not unique with respect to the underlying integration grid.; With $r_s \ge \Delta x$ we observe a unique one-interface solution irrespective of the initial conditions and on the partition of the integration scheme. 
The smooth character of the front indicates that the sensing radius generates a kind of surface tension between the moving $A$  and $B$ particles. Taking into account the particle numbers of adjacent cells when computing the diffusivities $D_A, D_B$ in \e{diff_lang} causes a smoothing  and avoidance of many drastic jumps in the probability densities.  

\bi{nonlocal-summary}b depicts the interface's maximal slope and how it depends on the sensing radius. The slope develops in nearly three steps with growing $r_s$. At small sensing radius it decreases rapidly. For moderate sensing sizes the slope  changes only weakly. When the sensing radius approaches half of the system size, the slope goes quickly to zero, indicating the transition to the  uniform distribution. We would like to point out that these result do not depend on the grid size -- several choices of $M$ show the same behavior.
\subsection{Hysteresis upon variation of coupling strength}
Finally, we discuss how the system with non-local interactions behaves for different values of the coupling coefficient $c$. In \bi{non-local2} we show how the structure of the stationary pdf changes with the parameter $c$ by global bifurcations. The following parameter continuation procedure was used. In \bi{non-local2}(a) we started with parameter $c=30$ from the initial conditions \e{initcond} with $p_B(x,0)=1-p_A(x,0)$ and $\delta=0.1$.
The parameter $c$ was decreased in steps of $1$, and the initial conditions were taken from the stationary pdfs of the previous parameter $c$ value. As can be seen, the number of interfaces becomes smaller as $c$ decreases and sequences of domains merge to larger areas of demixed states. The picture is nearly periodic and several collapses of domains happen for the same $c$-value. On the other hand, the parameter continuation in the forward direction, i.e. $c$ increases say from $4.2$, does not change the structure of stationary pdfs, which contains a single interface, as shown in \bi{non-local2}(b).  It means that the asymptotic state for higher $c$-values is multistable at least with respect to the various attractive states shown in the \bi{non-local2}(a) during the decrease of the $c$-values. This is further illustrated in \bi{non-local2}(c), which displays vertical cuts of backward and forward evolution of the densities for the indicated values of $c$. For $c=5$ stationary states obtained from either backward or forward continuation procedure are identical. In contrast, $c=10$ yields distinct stationary densities.
\begin{figure}[h!]
	\centering 
	\includegraphics[width=0.45\textwidth]{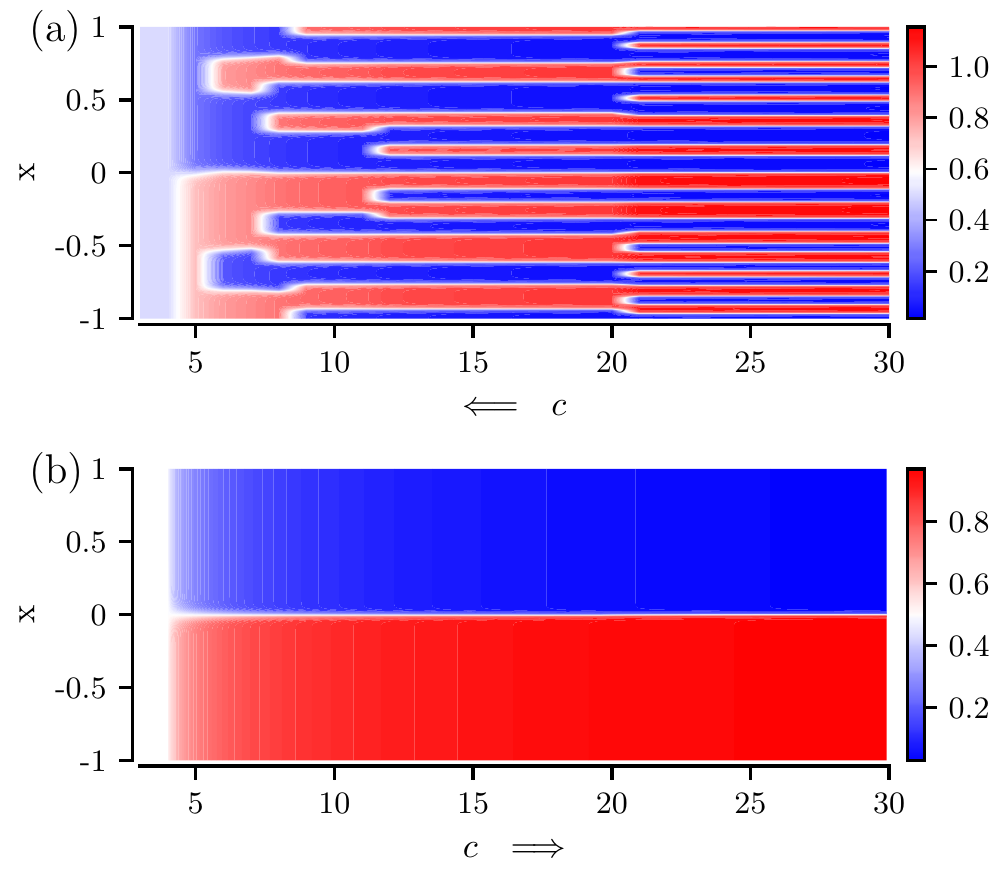}
	\includegraphics[width=0.45\textwidth]{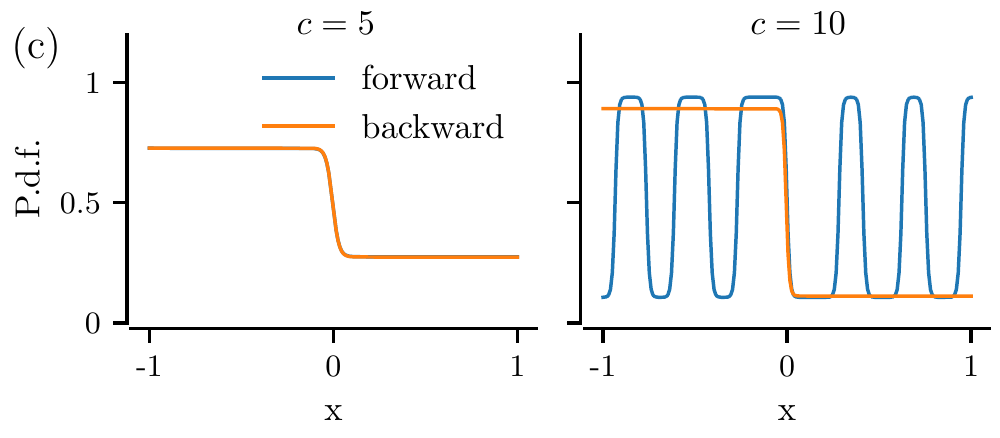}
	\caption{Parameter continuation of the stationary solutions for $N=200$ and $r_s=0.015$.	(a): Heat map of $p_B(x)$ vs $c$, for the backward continuation, when $c$ decreases from $c=30$ to $c=3$.
	(b): Heat map of $p_B(x)$ vs $c$, for the forward continuation, when $c$ increases from $c=4.1$ to $c=30$.
	(c): Stationary density $p_B(x)$ for the indicated values of $c$, for the backward and forward continuation.}
	\label{fig:non-local2}
\end{figure}

\section{Conclusions}
In summary, we considered a symmetric mixture of two species of stochastic micro-swimmers that move as overdamped Brownian particles in one dimension. We assumed that the effective diffusion coefficient of particles grows nonlinearly with the particle number of the  second species inside a sensing radius $r_s$. This assumption yields an instability inducing a spatial demixing of the two particle sorts above a critical strength of interaction. This demixing was demonstrated both  by particle simulations as well as by analytic inspection and numerical solution of the corresponding macroscopic (mean-field) Smoluchowski equations for the densities. The latter investigations revealed that the asymptotic solutions of the mean-field theory depend  on the size of the sensing radius. 

In case of a local definition with the radius of the order of an element of the integration grid, stable sequences of stepwise constant demixed domains with sharp interfaces were established as solutions of the coupled Smoluchowski equations  numerically and analytically. We can infer from the observed number of interfaces (jumps) that the local model in the absence of any mechanism for surface tension  possesses an ambiguous character; the slope of the interfaces in particle simulations depends on the chosen time step of integration; the spatial structure with multiple interfaces looks very different for different partitions of the underlying integration scheme for the corresponding Smoluchowski equation.

The simplest solutions for both densities are fully described by just two probability levels and are symmetric in the sense that they occupy equal domains of the entire interval.  We also found, however, asymmetric solutions in which $A$ and $B$ particles show different levels of demixing and, in order to describe the system, we need in total four levels of probability, with one pair of increase/decreased probability attained in two distinct domains of sizes $1\pm \Delta$. We characterized the (in)stability of these asymmetric solutions (only small values of asymmetry $\Delta$ lead to stable solutions) but also pointed out that the asymmetric solutions as well as the multi-interface solutions were not observable as long-time distributions in the particle simulations. This result was also in line with the Cahn-Hilliard potential  that was generally higher for the asymmetric states than for the symmetric ones, indicating that the asymmetric states are metastable and cannot survive in the Langevin simulations with finite-size fluctuations.

We then showed that a nonlocal definition of the diffusion coefficients in which 
we average over a neighborhood of the current grid point removed the ambiguity of asymptotic solutions of the Smoluchowski equations. Non-locality of the interaction  regularizes the stationary state. Arbitrary initial configurations end after a coalescence of domains in a unique stationary state, which in its front profile displays excellent agreement with the results of particle simulations.

With nonlocal sensing ($r_s > 0$ in the thermodynamic limit) a transition to a demixed state with asymptotically two domains have also been observed in two-dimensional particle simulations with periodic boundary conditions. As in the one-dimensional system, we obtain two distinct domains in which two distinct values of the probability are attained for the two species (e.g. increased for $A$ and decreased for $B$). As the boundary to the other domain is crossed, the two densities switch roles and jump to the respective other value (i.e. increased for $B$ and decreased for $A$). Both  regions are separated by a sharp interface; density values agree with the result of the analytic treatment given here. It remains an interesting task for future studies to explore novel features of demixing that are possible only in systems with higher spatial dimension \footnote{S.~Milster, private communication}.

\section*{Acknowledgement}
LSG thank Horst Malchow (Osnabr\"uck) and  Werner Ebeling (Rostock) for valuable comments on the manuscript.
ABN gratefully acknowledges the 	support of NVIDIA Corp. with the donation of the Tesla 	K40 GPU used for this research.

\appendix

\section{Eigenvalues of the steady states and interpretation of multiplicative noise}
\label{sec:eigenvalues_and_strato}
The connection between the Langevin equations with multiplicative noise \cite{hanggi1982stochastic,risken1996fokker} and the corresponding Smoluchowski equation was formulated in the main text for the Ito interpretation of the corresponding differential equation. For completeness, we give here the details of the  eigenvalue analysis for this interpretation in the case of a local coupling (the generalization to nonlocal coupling is straightforward). In addition, we repeat the stability analysis for the case of the Stratonovich interpretation of the stochastic differential equations and demonstrate that in this case the homogeneous state is always stable with a quadratic nonlinearity.

The Smoluchowski equations in the Ito interpretation of the Langevin equations read \cite{risken1996fokker}
\ba
\partial_t \, p_A \,= \,\partial_x^2 \{ f(p_B(x))\, p_A(x)\}\nonumber \\
\partial_t \, p_B \,= \,\partial_x^2 \{ f(p_A(x))\, p_B(x) \}
\ea
where we used for the ease of notation the abbreviation $f(p)=1+cp^2$. Let us suppose
that we have found steady states $p_A^0$ and $p_B^0$, i.e. constant values in a certain domain. This might be the homogeneous solution, the symmetric solution from \se{symm} in two domains of identical  size $1$, or the solutions in the asymmetric domains of size $1+\Delta$ and $1-\Delta$ from \se{asymm}. In any case, we add small spatio-temporal perturbations to these steady-state solutions,  $p_A(x,t)=p_A^0 + \delta p_A(x,t)$ and $p_B(x,t)=p_B^0 + \delta p_B(x,t)$ with $\delta p(x,t) \propto \exp(\lambda t +{\rm i} k x)$ and linearize the problem with respect to small $\delta p_{A,B}$. In this way, one obtains
\ba
\partial_t \, \delta p_A\,=\, - \,f(p_B^0) k^2 \delta p_A\,-\,p_A^0 f^\prime(p_B^0) k^2 \delta p_B \nonumber \\
\partial_t \, \delta p_B\,=\, - \,p_B^0 f^\prime(p_A^0) k^2 \delta p_A \,- \,f(p_A^0) k^2 \delta p_B\,
\label{eq:eigen_ito}
\ea
where $f^\prime(p)$ stands for the derivative with respect to the argument of the function. From \e{eigen_ito} we obtain the eigenvalues as nontrivial solutions $\lambda \ne 0 $ of a quadratic equation:
\ba
\frac{\lambda_{1,2}}{k^2}&=&-\left(1+\frac{c}{2}(p_{A,0}^2+p_{B,0}^2)\right) \pm \nonumber \\
&& \frac{c}{2}\sqrt{p_{A,0}^4+14p_{A,0}^2p_{B,0}^2+p_{B,0}^4}.
\label{eq:eigenvalues_general} 
\ea
For the uniform densities for the entire domain ($p_{A,0}=p_{B,0}=1/2$), we obtain 
\be
\frac{\lambda_{1,2}}{k^2}=-1+\begin{cases}\frac14 c\\ -\frac34 c \end{cases}
\ee
which corresponds in its first (larger) solution to \e{disp_local} in the main text and shows in particular that one of the eigenvalues becomes positive for $c>c_{crit}$ and the uniform solution looses its stability. For such supercritical values of the coupling coefficient, we have found the alternative symmetric solutions, say, $p_{A,h}$ and $p_{B,l}$ attained in one domain of size 1 and symmetrical solutions $p_{A,l}$ and $p_{B,h}$ in an identical domain, where the values are given by (repeating \e{sol_symm} from the main text)
\ba
p_{A,h}&=&p_{B,h}=\frac12+\sqrt{\frac14-\frac1{c}} \\
p_{A,l}&=&p_{B,l}=\frac12-\sqrt{\frac14-\frac1{c}}
\ea
The eigenvalues in this case read
\be
\frac{\lambda_{1,2}}{k^2}=-\frac{c}2\pm \sqrt{\frac{c^2}4+4-c}
\ee
and it is not hard to see that for $c>c_{crit}=4$, both eigenvalues are negative, i.e. the symmetric solutions that show a demixing are stable against small fluctuations. 
\begin{figure}[h!] 
\centering
\includegraphics[width=0.45\textwidth]{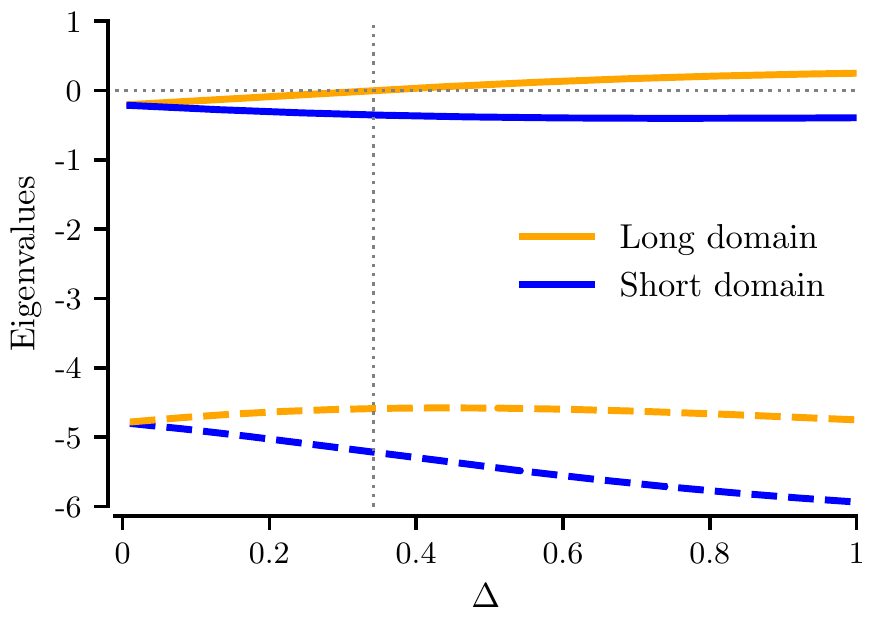}
\caption{Eigenvalues for asymmetric solutions vs. excess in domain size $\Delta$. 
This plot complements \bi{asymm_map} and reveals that these states loose stability for $\Delta\ge 0.34$ where one of the eigenvalues becomes positive. The shown lines have been computed using \e{eigenvalues_general}  with  $p_{A,0}=p_{A,l}, p_{B,0}=p_{B,h}$ and  $p_{A,0}=p_{A,h}, p_{B,0}=p_{B,l}$ obtained as numerical solutions from  \e{normi_as} and \e{stat_smolu_as}.
}
	\label{fig:eigenvalues}
\end{figure}

Finally, if we assume asymmetric domains of size $1+\Delta$ and $1-\Delta$, we were not able to find explicit solutions but had to resort to the numerical solution of \e{normi_as} and \e{stat_smolu_as}. We would like to mention, however, that the four nonlinear equations are partly linear in some of the variables and this fact can be used to reduce the problem to \emph{one} (highly nonlinear) equation for a remaining variable (say, $p_B,l$). Because this resulting equation is very lengthy and not insightful,  we abstain from presenting it here. We have inspected the equation graphically and made sure that (at least for our standard value $c=5$) the discussed solution for the four values $p_{A,l},p_{A,h},p_{B,l}, p_{B,h}$ is the unique solution of the problem (apart from the trivial second solution obtained by swapping the indices $A$ and $B$). If we insert the pairs of values into \e{eigenvalues_general}, we obtain the eigenvalues shown in \bi{eigenvalues}.

Let us return to the stability of the uniform distribution from a more general perspective and also discuss what happens if we switch from the Ito interpretation to the Statonovich interpretation.

Considering the linear system in \e{eigen_ito}, we see quickly that a saddle-node bifurcation  (one real eigenvalue changing sign) occurs if 
\be
 p_A^0 p_B^0 f^\prime(p_A^0)f^\prime (p_B^0)-f(p_A^0)f(p_B^0)\geq 0\,.
\ee
For the symmetric homogeneous distribution this reduces to the simpler condition  $p^0 f^\prime(p^0) -f(p^0) \geq 0$, which means $c q (p^0)^{q}-(1+c(p^0)^q)\geq 0$. A minimal requirement is $q > 1$. For $q=2$ and $p_0=1/2$ the instability occurs at $c=4$ as calculated above and in the main text, see \e{crit}.

Let us now turn to the Stratonovich calculus, i.e. interpret the stochastic differential equations in the sense of Stratonovich \cite{risken1996fokker}. This yields the Smoluchowski equation 
\ba
\partial_t p_A = \partial_x\sqrt{f(p_B(x))} \partial_x\sqrt{f(p_B(x)}\, p_A(x),\nonumber \\
\partial_t p_B = \partial_x\sqrt{f(p_A(x))} \partial_x\sqrt{f(p_A(x)}\, p_B(x).
\ea
We proceed as above and assume small wave-like perturbations around the fixed homogeneous states $p_A^0$ and $p_B^0$, i.e. $p_A(x,t)=p_A^0 + \delta p_A(x,t)$ and $p_B(x)=p_B^0 + \delta p_B(x,t)$ with  $\delta p(x,t) \propto \exp(\lambda t +{\rm i} k x)$ and obtain
\ba
&&\lambda \, \delta p_A= - f(p_B^0) k^2 \,\delta p_A - \frac{1}{2} p_A^0 f^\prime(p_B^0) k^2\,\delta p_B\,,\nonumber \\ 
&&\lambda \, \delta p_B=  - \frac{1}{2} p_B^0 f^\prime(p_A^0) k^2 \,\delta p_A - f(p_A^0) k^2 \,\delta p_B\,.
\ea
The difference to the Ito case is the appearance of factors  $1/2$ in front of the second or first term, respectively.  In consequence, one obtains for the two eigenvalues again a quadratic equation and the instability in the Stratonovich case appears  if 
\be
\frac{1}{4} p_A^0 p_B^0 f^\prime(p_A^0)f^\prime (p_B^0)-f(p_A^0)f(p_B^0)\ge 0\,.
\ee
We now inspect the condition for the homogeneous case  with $p_A^0\,=\,p_B^0\,=1/2$ and our nonlinearity $f(p)=1+c p^q$. Generally, for the homogeneous state the instability takes place if $p^0 \,f^\prime(p^0) /2\, \geq \,f(p^0)S$. Remarkably, we find that for the nonlinearity parameter $q=2$ no bifurcation with respect to $c$ occurs. For the Stratonovich case one obtains that instabilities appear if \begin{equation}
\bracket{\frac{1}{2}\,q\,-\,1}\,c\,(p^0)^q\,\geq \, 1.
\label{eq:eigen_strato}
\end{equation}
which requires a stronger nonlinearity with $q\geq 3$. The bifurcation is then attained for 
\be
c \geq 2^{q+1}/\bracket{q\,-\,2}.
\ee 

\section{Numerical solution of Smoluchowski equations}

The original equations are
\begin{eqnarray}
&&\partial_t p_A (x,t)=\partial_x^2 \left\{ f(p_B(x))p_A(x)  \right\}, \nonumber \\
&&\partial_t p_B (x,t)=\partial_x^2 \left\{ f(p_A(x))p_B(x)  \right\}.
\label{eq:local}
\end{eqnarray}
For the nonlocal case, nonlinear function $f[\cdot]$ in \e{local} was applied to the corresponding averaged probability density,
\begin{equation}
\widetilde{p}_{A,B}(x,t)=\frac{1}{2r_s}\int_{-r_s}^{r_s} p_{A,B}(x+y)dy.
\end{equation}
Discretisation in the spatial variable, $x\to x_n=-1+(n-1)\Delta x$, $\Delta x=2/(N-1)$, $n=1,...,N$, gives the system of $2N$ Ode's,
\begin{eqnarray}
&&\dot{p}_{A,n}=\frac{1}{h^2} \left[ P_{n+1}-2P_n + P_{n-1} \right],\nonumber\\
&&\dot{p}_{B,n}=\frac{1}{h^2} \left[ Q_{n+1}-2Q_n + Q_{n-1} \right], \nonumber\\
&& P_n = f\left(p_{B,n}\right) p_{A,n}, \quad Q_n = f\left(p_{A,n}\right) p_{B,n}.
\label{eq:discrete}
\end{eqnarray}
These equations were solved using MATLAB solver ode15s for stiff
Ode's. The no-flux boundary conditions, $\partial_x p_{A,B}|_{x=\pm1}=0$, were used in all figures showing solutions of the Smoluchowski equation.

In numerical implementation the sensing radius is given by \e{rs-disc} and we calculated the moving averages:
$$
\widetilde{p}_n=\frac{1}{2s-1} \sum_{m=1-s}^{s-1} p_{n+m}.
$$
These were then
used in calculations of the $P_n$ and $Q_n$ in \e{discrete}:
$P_n=f[\widetilde{p}_{B,n}]p_{A,n}, \quad Q_n=f[\widetilde{p}_{A,n}]
p_{B,n}.$

Numerical integration of Ode's was carried out with an absolute
tolerance of $10^{-9}$. The approach to a stationary solution was
controlled by calculating the maximum difference over $x$ between
probability densities at two consecutive integration intervals, i.e.
$$\varepsilon = \max {|p_{A,B}(x,t+\Delta t)-p_{A,B}(x,t)|},$$
and stopped when $\varepsilon < 10^{-8}$.

\end{document}